\documentclass[twocolumn,prl,aps,superscriptaddress,showpacs]{revtex4-1}
\usepackage[latin9]{inputenc}
\usepackage{amsmath}
\usepackage{amssymb}
\usepackage{graphicx}

\makeatletter

\DeclareTextSymbolDefault{\textquotedbl}{T1}

\usepackage{amsfonts}
\usepackage{color}
\usepackage{soul}

\makeatletter
\newcommand{\subalign}[1]{%
  \vcenter{%
    \Let@ \restore@math@cr \default@tag
    \baselineskip\fontdimen10 \scriptfont\tw@
    \advance\baselineskip\fontdimen12 \scriptfont\tw@
    \lineskip\thr@@\fontdimen8 \scriptfont\thr@@
    \lineskiplimit\lineskip
    \ialign{\hfil$\m@th\scriptstyle##$&$\m@th\scriptstyle{}##$\crcr
      #1\crcr
    }%
  }
}
\makeatother
\begin{document}


\title{Error suppression in adiabatic quantum computing with qubit ensembles}

\author{Naeimeh Mohseni}
\affiliation{State Key Laboratory of Precision Spectroscopy, School of Physical and Material Sciences,East China Normal University, Shanghai 200062, China}
\affiliation {Max-Planck-Institut f{\"u}r die Physik des Lichts, Staudtstrasse 2, 91058 Erlangen, Germany}

\affiliation{ Department of Physics, Institute for Advanced Studies in Basic Sciences(IASBS), Zanjan 45137-66731, Iran}

\author{Marek Narozniak}
\affiliation{New York University Shanghai, 1555 Century Ave, Pudong, Shanghai 200122, China}
\affiliation{Department of Physics, New York University, New York, NY 10003, USA}

\author{Alexey N. Pyrkov}
\affiliation{Institute of Problems of Chemical Physics RAS, Acad. Semenov av., 1, Chernogolovka, 142432, Russia}
\author{Valentin Ivannikov}
\affiliation{New York University Shanghai, 1555 Century Ave, Pudong, Shanghai 200122, China}
\affiliation{State Key Laboratory of Precision Spectroscopy, School of Physical and Material Sciences,East China Normal University, Shanghai 200062, China}

\author{Jonathan P. Dowling}
\affiliation{Hearne Institute for Theoretical Physics, Department of Physics and Astronomy,
Louisiana State University, Baton Rouge, Louisiana 70803, USA}
\affiliation{NYU-ECNU Institute of Physics at NYU Shanghai, 3663 Zhongshan Road North, Shanghai 200062, China}

\author{Tim Byrnes}
\email{Corresponding author: tim.byrnes@nyu.edu}
\affiliation{New York University Shanghai, 1555 Century Ave, Pudong, Shanghai 200122, China}
\affiliation{State Key Laboratory of Precision Spectroscopy, School of Physical and Material Sciences,East China Normal University, Shanghai 200062, China}

\affiliation{NYU-ECNU Institute of Physics at NYU Shanghai, 3663 Zhongshan Road North, Shanghai 200062, China}
\affiliation{National Institute of Informatics, 2-1-2 Hitotsubashi, Chiyoda-ku, Tokyo 101-8430, Japan}
\affiliation{Department of Physics, New York University, New York, NY 10003, USA}

\date{\today}


\begin{abstract}
Incorporating protection against quantum errors into adiabatic quantum computing (AQC) is an important task due to the inevitable presence of decoherence. Here we investigate an error-protected encoding of the AQC Hamiltonian, where qubit ensembles are used in place of qubits. Our Hamiltonian only involves total spin operators of the ensembles, offering a simpler route towards error-corrected quantum computing. Our scheme is particularly suited to neutral atomic gases where it is possible to realize large ensemble sizes and produce ensemble-ensemble entanglement. We identify a critical ensemble size $N_{\mathrm{c}}$ where the nature of the first excited state becomes a single particle perturbation of the ground state, and the gap energy is predictable by mean-field theory. For ensemble sizes larger than $N_{\mathrm{c}}$, the ground state becomes protected due to the presence of logically equivalent states and the AQC performance improves with $N$, as long as the decoherence rate is sufficiently low.
\end{abstract}
 
\maketitle

\section*{INTRODUCTION}

Adiabatic quantum computing (AQC) is an alternative approach to traditional gate-based quantum computing where quantum adiabatic evolution is performed in order to achieve a computation \cite{farhi2000quantum,farhi2001quantum,RevModPhys.80.1061,RevModPhys.90.015002}.  In the scheme, the aim is to find the ground state of a Hamiltonian $ H_Z $ which encodes the problem to be solved and can be considered an instance of quantum annealing \cite{finnila1994quantum,kadowaki1998quantum,santoro2002theory,brooke1999quantum}. In addition, an initial Hamiltonian $ H_X $, which does not commute with the problem Hamiltonian, is prepared such that the ground state is known.  For example, a common choice of these Hamiltonians are
\begin{align}
H_Z^{\text{(qubit)}} & = \sum_{i=1}^M \sum_{j=1}^M J_{ij} \sigma_i^z \sigma_j^z +  \sum_{i=1}^M K_i \sigma_i^z,  \label{hzham} \\
H_X^{\text{(qubit)}}  & = - \sum_{i=1}^M \sigma_i^x, \label{hxham}
\end{align}
where $ \sigma_i^{x,z} $ are Pauli matrices on site $ i $, and $ J_{ij} $ and $ K_i $ are coefficients which determine the problem to be solved, and there are $ M $ qubits.  The form of (\ref{hzham}) directly encodes a wide variety of optimization problems, for example, MAX-2-SAT and MAXCUT which are NP-complete problems.  It then follows that any other problem in NP can be mapped to it in polynomial time \cite{mezard1987spin,Papadimitriou1995,Garey1979}. AQC then proceeds by preparing the initial state of the quantum computer in the ground state of $ H_X $, then applying the time-varying Hamiltonian
\begin{align}
H = (1-\lambda(t) ) H_X + \lambda(t) H_Z, 
\label{aqcham}
\end{align}
where  $ \lambda(t) $ is a time-varying parameter that is swept from $ 0 $ to $ 1 $.  Intense investigation into the performance of AQC has been performed since its original introduction, demonstrating its performance for various problems \cite{van2001powerful,roland2002quantum,hogg2003adiabatic,amin2008effect} and characterizing  the effects of decoherence \cite{childs2001robustness,sarandy2005adiabatic,roland2005noise,lidar2008towards,ashhab2006decoherence,amin2009decoherence,deng2013decoherence}.

In the AQC framework, the speed of the computation is given by how fast the adiabatic sweep is performed. To maintain adiabaticity, one must perform the sweep sufficiently slowly, such that the system remains in the ground state throughout the evolution. The sweep time required to maintain adiabaticity is known to be proportional to a negative power of the minimum energy gap of the Hamiltonian (\ref{aqcham}),  where the power depends upon the annealing schedule $ \lambda(t) $ and gap structure \cite{farhi2001quantum,roland2002quantum,amin2008effect,aharonov2003adiabatic,schaller2006general,jansen2007bounds,lidar2009adiabatic}. 
One of the attractive features of AQC is that time-sequenced gates do not need to be applied, but it  is nevertheless known to be equivalent to the gate-based quantum computation \cite{mizel2007simple,aharonov2008adiabatic,roland2002quantum,wei2006modified,PhysRevA.65.062310,jiang2018quantum,PhysRevLett.108.130501}.   Numerous theoretical analyses  and experimental demonstrations have been performed both at small 
\cite{steffen2003experimental,mitra2005experimental,johnson2011quantum,boixo2013experimental,barends2016digitized} and larger scale \cite{shin2014quantum, PhysRevA.90.042317, PhysRevA.94.062106,bauer2015entanglement,boixo2014evidence,ronnow2014defining,albash2015reexamining,heim2015quantum,preskill18}. 
One of the outstanding problems for AQC is to fully understand the performance and effectively combat decoherence in AQC such that it can be applied to real-world combinatorial problems \cite{RevModPhys.90.015002,young2013error,sarovar2013error}.

In this paper, we investigate a variant of the AQC Hamiltonians (\ref{hzham}) and (\ref{hxham}) where ensembles of qubits are used to encode the optimization problem, instead of genuine qubits.  Specifically, we study the Hamiltonians
\begin{align}
H_Z & = \frac{1}{N} \sum_{i=1}^M \sum_{j=1}^M   J_{ij} S_i^z S_j^z +   \sum_{i=1}^M  K_i S_i^z,   \label{hzhamens}  \\
H_X & = -  \sum_{i=1}^M  S_i^x, \label{hxhamens}
\end{align}
where $ S_i^{x,z} = \sum_{n=1}^N \sigma_{i,n}^{x,z} $ are total spin operators for an ensemble consisting of $ N $ qubits.  In this case, $ M $ is the number of ensembles.  Here, $ \sigma_{i,n} $ denotes the Pauli operator for the $ n $th qubit within the $i$th ensemble.  The matrices $ J_{ij} $ and $ K_i $ are the same as that in (\ref{hzham}) and we take $ J_{ij} = J_{ji} $ and $ J_{ii} = 0 $. AQC then proceeds in the same way as described in Eq.  (\ref{aqcham}).  Each of the ensembles is initially prepared in a fully polarized state of  $ \langle S_i^{x}  \rangle = N $ and adiabatically evolved to the ground state of $H_Z $.  The aim will be to investigate whether the ensemble version of the Hamiltonian can be used in place of the qubit Hamiltonian, such that the ground state configuration of  (\ref{hzham}) is found using (\ref{hzhamens}) and (\ref{hxhamens}). We characterize the nature of the ground and excited states of the ensemble Hamiltonian and assess the performance of AQC in comparison to the original qubit problem. 

The Hamiltonians (\ref{hzhamens}) and (\ref{hxhamens}) can be considered an error-suppressing encoding of the original AQC Hamiltonians  (\ref{hzham}) and (\ref{hxham}), respectively.  The use of an ensemble duplicates the quantum information since the $ N $ qubits within an ensemble are in the same state at the start and at the end of the adiabatic evolution.  Such error-suppression strategies have been already investigated in the context of AQC. Jordan, Farhi, and Shor \cite{jordan2006error} introduced an encoding capable of detecting the presence of single-qubit errors, which are suppressed by an additional energy penalty term in the total Hamiltonian. Pudenz, Lidar and co-workers \cite{pudenz2014error}, introduced a scheme known as quantum annealing correction (QAC), where a repetition code is used to encode a logical qubit and a majority vote is used to decode the logical operations. This, combined with the addition of energy penalty terms to enforce alignment within each ensemble have shown that the scheme is effective at achieving error-suppression \cite{PhysRevLett.116.220501,PhysRevA.95.022308,mishra2016performance,vinci2015quantum,pearson2019analog}. In particular, the two-qubit interaction term in (\ref{hzham}) is encoded by repeating the interactions $ N $ times in a pair-wise fashion between the logical qubit ensembles, motivated by the chimera qubit connectivity of the D-wave quantum computers.  An alternative scheme called nested QAC has also been studied by Vinci, Lidar and co-workers, where the qubit terms are mapped with an all-to-all interaction, similar to the one we consider in (\ref{hzhamens}) \cite{vinci2016nested}. A similar all-to-all encoding was considered by Venturelli, Smelyanskiy and co-workers \cite{venturelli2015quantum}. 
 A minor embedding is then performed on the encoded Hamiltonian to match it to the chimera graph topology  \cite{vinci2018scalable}. Matsuura and co-workers showed through mean-field analysis that the order of the phase transition is modified through the mapping in the ferromagnetic and antiferromagnetic Ising models \cite{PhysRevA.99.062307}. Young, Sarovar, and Blume-Kohout \cite{young2013error,sarovar2013error} showed that such energy penalty approaches could be part of a unified theory with dynamical decoupling.

\begin{figure}
\centering
\includegraphics[width=1\linewidth]{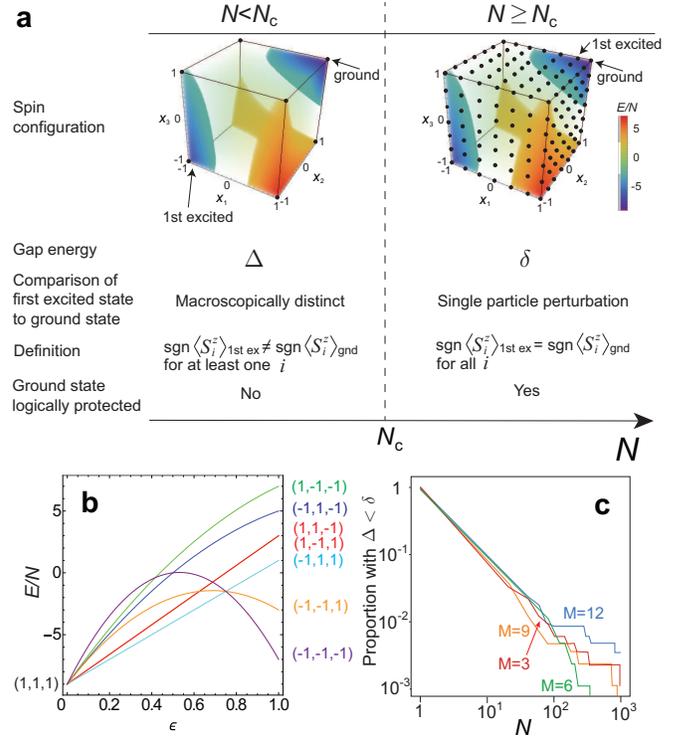} 
\caption{Properties of the problem Hamiltonian $ H_Z $.  (a) Summary of the two regimes for the ensemble version of the problem Hamiltonian $ H_Z $.  The cube shows the energy landscape of a typical instance of (\ref{hzhamens})  with rescaled variables $ x_i = \langle S_i^z \rangle/N $.  Dots indicate allowed states for $ N =1 $ (left) and $ N =6 $ (right).  Only states on the visible faces are shown for clarity.  
(b) Energy variation for the problem instance as Fig. \ref{fig1}(a) along a linear trajectory from the ground state $ x_1 = x_2=x_3 =1 $ to other hypercube corners specified by $ (x_1^{(f)}, x_2^{(f)}, x_3^{(f)}) $.  The trajectory is defined by $x_i = 1-(1-x_i^{(f)}) \varepsilon $, where $ \varepsilon = 0 $ is the ground state and $ \varepsilon = 1 $ is another hypercube corner. Parameters used are $ M = 3 $, $ J_{12} = -2,  J_{13} = -1, J_{23}=-1, K_{1} = 1,  K_{2} = 0,  K_{3} = -2 $.  (c) Proportion of 800 randomly generated problem instances with $ \Delta < \delta $ (i.e. $ N < N_{\mathrm{c}} $) as a function of the ensemble size $ N $. Elements of $ J_{ij} $ and $ K_i $ are taken randomly in the interval $ \left[ -1,1 \right] $ with a uniform distribution. 
 \label{fig1}}
\end{figure}

\section*{RESULTS }

\subsection*{Properties of the problem Hamiltonian $H_Z$ }

We first examine the properties of the ensemble version of the (classical) problem Hamiltonian (\ref{hzhamens}).  The typical energy landscape of the Hamiltonian $ H_Z $  is shown in Fig. \ref{fig1}(a).  The axes are plotted with rescaled spin variables $ x_i = \langle S_i^z \rangle /N $. For the corners of the hypercube $ x_i = \pm 1 $, the energy eigenvalues of the ensemble Hamiltonian  $ H_Z $ reduces to that of the qubit version  (\ref{hzham}) up to an overall scaling factor of $ N $. This is a general result which is true by virtue of the structure of the ensemble and qubit problem Hamiltonians $ H_Z $ (see Appendix A).  The primary difference between (\ref{hzhamens}) and (\ref{hzham}) is then that the ensemble version can take a discrete set of intermediate values of $ x_i $ between the $ \pm 1 $ values. 

In Fig. \ref{fig1}(b) we show the variation of the energy starting from the ground state to the remaining hypercube corners. The variation always follows a quadratic form with an initially positive slope.  This can be shown to be generally true following from the quadratic form of (\ref{hzhamens}) (see Appendix A).  This fact can be used to show that for any point along a trajectory connecting the ground state to another hypercube corner has energies greater than the ground state  (see Appendix A).  
From the above structure of the energy landscape, one can deduce that the ground states of the Hamiltonians (\ref{hzham}) and (\ref{hzhamens}) have logically equivalent spin configurations.  We define the logically equivalent states of the qubit and ensemble systems according to 
\begin{align}
\text{sgn} [ \langle \sigma_i^z \rangle^{\text{(qubit)}} ]  = \text{sgn} [ \langle S_i^z \rangle^{\text{(ens)}} ] ,
\label{equivalent}
\end{align}
for all $ i $. This is also known as a majority vote encoding of the ensemble to give the logical state, and has been considered in other error mitigation schemes for AQC \cite{pudenz2014error,vinci2016nested,young2013error}.  Thus, finding the ground state spin configuration of (\ref{hzhamens}) gives the same information as (\ref{hzham}).

In AQC, one of the parameters which plays a central role is the gap energy, i.e. the energy between the ground and first excited state. The simple structure of the Hamiltonian (\ref{hzhamens}) allows us to deduce that for a given $ N $, there are two different regimes for the gap energy.  For the particular example shown in Fig. \ref{fig1}(a), we see that there are two hypercube corner states $ (x_1,x_2,x_3) = (1,1,1) $ and $  (-1,-1,-1)$ with relatively similar energies of $ \epsilon_0 $ and $ \epsilon_1 $, respectively.  For the qubit case (\ref{hzham}) the difference between the two lowest energy hypercube corners is the gap energy.  In the ensemble case, the energy difference between these two hypercube corners is
\begin{align}
\Delta = N (\epsilon_1 -\epsilon_0 ),
\label{Deltaform}
\end{align}
since for extremal values $ |x_i| =  1 $ the energies are related by a factor of $ N $. 

If $ N $ is sufficiently small, $ \Delta $ remains the gap energy for the ensemble case.  However, for  large enough $ N $ this becomes less and less likely,   and the first excited state is a single qubit spin-flip of the ground state.  Specifically, we have the state such that on the $k$th ensemble  $ S_k^z = \pm (N-2) $, and the remaining ensembles $ S_i^z = \pm N,  \forall i \ne k $. The energy gap for this single qubit flip state is 
\begin{align}
\delta = \min_{k} \left[ - 2 \sigma_k \left(  K_k + 2 \sum_{j} J_{kj} \sigma_j \right) \right] ,
\label{deltaexpression}
\end{align}
where $ \sigma_i = \langle \sigma_i^z \rangle^{\text{(qubit)}}_{\text{gnd}} = \text{sgn} ( \langle S_i^z \rangle^{\text{(ens)}}_{\text{gnd}} ) $, and expectation values are taken with respect to the ground state.  The minimum function picks the smallest value from the range $ k \in [1,M] $. 

Whether $ \Delta $ or $ \delta $ is the gap energy depends upon $N $ and the particular parameter choice of $ J_{ij} $ and $ K_i $ made.   As $ N $ is increased, at some point there will always be a crossover such that the gap is $ \delta $, since (\ref{Deltaform})  is proportional to $ N $ while (\ref{deltaexpression}) has no dependence on $ N $.  Let us call $ N_{\mathrm{c}} $ the smallest value of $ N $ such that $ \Delta > \delta $.  For a given problem instance, we then define two regions of $ N $,  according to whether it is larger or smaller than $ N_{\mathrm{c}} $. The two regimes and their implications summarized in Fig. \ref{fig1}(a).  
In Fig. \ref{fig1}(c) we show the proportion of problems satisfying $ \Delta < \delta $ (i.e. $ N < N_{\mathrm{c}} $) for randomly generated $ J_{ij} $ and $ K_i $.  We see that the proportion decreases as $ \propto 1/N $, which is consistent with the linear scaling of $ \Delta $.  We note that in the case of atomic ensembles $N $ can be quite large (e.g. $ 10^3 $ to $ 10^{11} $) \cite{julsgaard2001experimental,fadel2017}, which suggests that for realistic ensemble sizes most of problem instances will be in the regime $ N \ge N_{\mathrm{c}} $.

\begin{figure*}
\begin{center}
\includegraphics[width=0.9\linewidth]{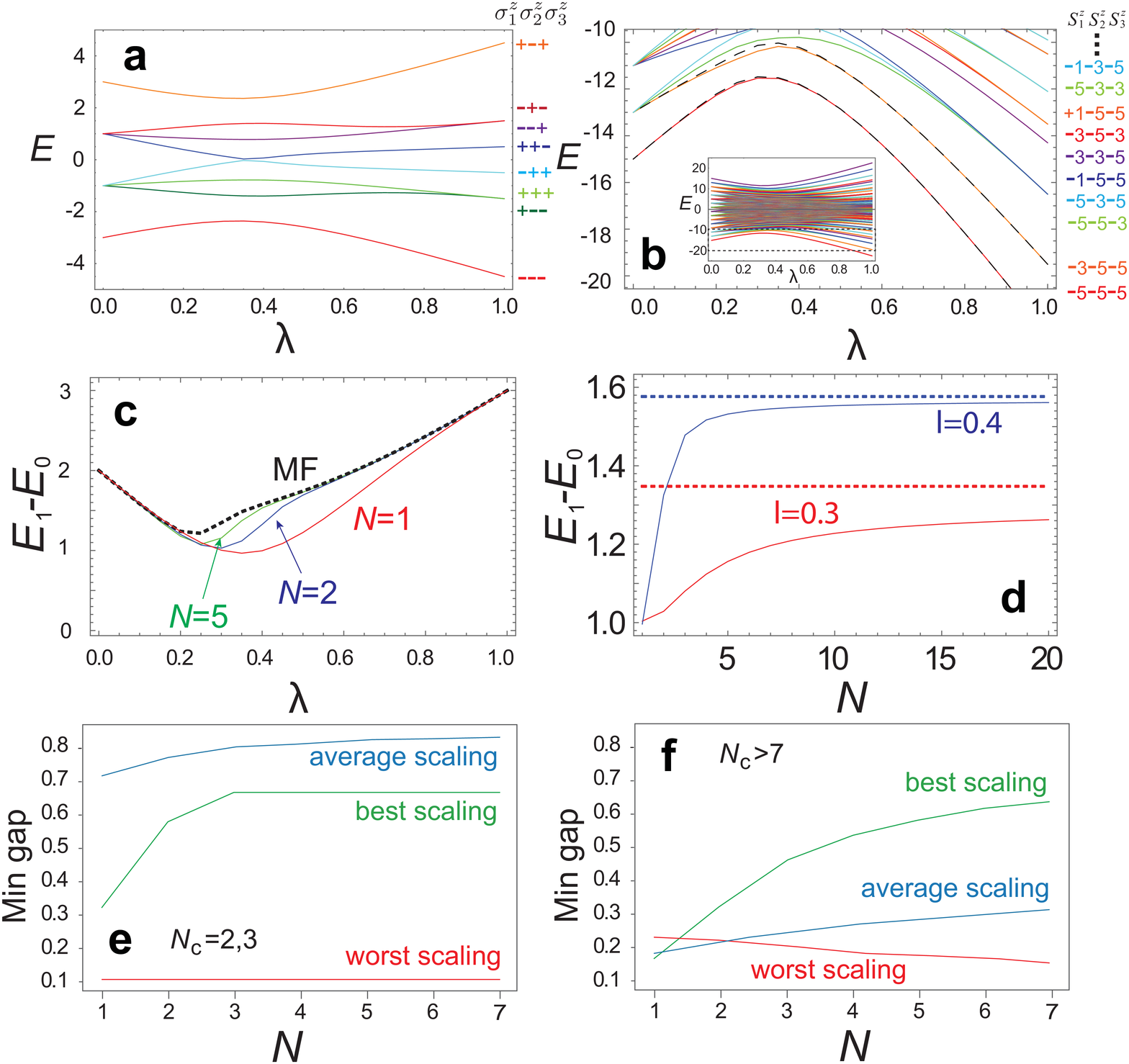} 
\end{center}
\caption{Energy spectrum and gap energies of the adiabatic quantum computing Hamiltonian. Spectrum of  (\ref{aqcham}) with $ M =3 $ with  (a)  $ N = 1 $ and  (b) $ N = 5 $ for parameters  $J_{12}=-0.5, J_{13}=0, J_{23}=-1, K_1=0.5, K_2=0, K_3=1 $.  The mean-field approximation for the $N = 5 $ is shown as the dashed lines for the ground and first excited state. (c) The gap energy for the ensemble qubit numbers as shown.  (d) Scaling of the gap energy with $N $ for two values of $ \lambda $ as shown.  The mean-field (MF) approximation is shown as the dotted line in (c) and (d). (e) The minimum gap versus $ N$ for 60 random instances of problems with $ N_{\mathrm{c}} = 3 $. The average minimum gap, as well as the largest (best) and smallest (worst) instances are shown.  (f) Same as (e) for problems with $ N_{\mathrm{c}} > 7 $.  
\label{fig2}} 
\end{figure*}

\subsection*{Spectrum of the AQC Hamiltonian}
So far we have only examined the classical limit of $ \lambda = 1 $. The overall speed of the AQC will be dependent upon the minimum gap energy with the off-diagonal term (\ref{hxhamens}) present. To illustrate the effect of intermediate $ \lambda $, we compare the eigenvalue spectrum of the Hamiltonian (\ref{aqcham}) for the standard qubit case and the ensemble  case with $ N = 5 $ for the same $J_{ij} $ and $K_i $ parameters in Fig. \ref{fig2}(a)(b). 
Due to symmetry under particle interchange on each ensemble, the Hilbert space can be reduced to the symmetric subspace, reducing the dimensionality from $ 2^{NM} $ to $ (N+1)^M $. The most noticeable difference is the larger number of states when ensembles are used
(Fig. \ref{fig2}(b) inset).  Despite the larger number of states, plotted on the same energy scale, a non-diminishing gap between ground and excited state maintained for the ensemble case (Fig. \ref{fig2}(b) main figure).  This occurs due to the larger energy scale of the ensemble Hamiltonian by a factor of $ N $, which at least partially compensates for the larger number of states.  
 Many of these additional states are logically equivalent states in the sense of (\ref{equivalent}).  For example, we label the states at $ \lambda = 1 $ in terms of the eigenstates $ S^z_i $. In the qubit version the two lowest states have a spin configuration of $ ( \sigma_1^z, \sigma_2^z,\sigma_3^z) = (-1,-1,-1) $ and $ ( +1, -1, -1) $ respectively.   In the ensemble version with $ N = 5 $, of the shown states, the lowest 7 states  are all logically equivalent to  $(-1,-1,-1) $ in terms of (\ref{equivalent}).  Such logically equivalent states provide protection against error since they occur with energies in the vicinity of the ground state, and act as as a ``buffer'' before logical errors are induced \cite{young2013error, pudenz2014error}.

Our aim in the AQC will be to keep the adiabatic evolution in the ground state of the ensemble system. Obtaining the ground state and first excited state for the ensemble system in general is a numerically intensive task involving a diagonalization within a Hilbert space of dimension $ (N+1)^M $.  To see the behavior for large ensemble sizes, it is desirable to have an approximate method of estimating the gap energy that does not require full diagonalization.  Mean-field theory provides an accurate estimate of physical quantities for large spin systems. The ensemble nature of the Hamiltonian allows us to extract energies with increasing accuracy particularly for large $ N $.   We use a mean-field ansatz wavefunction of the form
\begin{align}
| \Psi_{\text{MF}}^{(0)} \rangle = \prod_{i=1}^M | 0, \theta_i \rangle_i,
\label{ground}
\end{align}
where we define a Fock state of $ N $ spins all aligned in the same direction as
\begin{align}
| 0, \theta \rangle_i =  \prod_{n=1}^N \left( \cos \frac{\theta}{2} |0 \rangle_{ni} + \sin \frac{\theta}{2} |1 \rangle_{ni} \right)  ,
\end{align}
which is the maximal positive eigenstate of the rotated spin operator $ \tilde{S}^z = \sin \theta S^x + \cos \theta S^z $. We note that a similar mean-field ansatz was used in past works to analyze the $N = 1 $ AQC Hamiltonian ground state \cite{smolin2014classical,shin2014quantum,PhysRevA.94.062106,albash2015reexamining}. 
We apply the mean-field ansatz by performing a self-consistent procedure to obtain the parameters $ \theta_i $.  This is equivalent to taking expectation values of the Hamiltonian (\ref{aqcham}) 
with respect to (\ref{ground}) and optimizing for $ \theta_i $  (see Appendix B).   From the discussion relating to the logically equivalent states, a suitable mean-field ansatz for the first excited state consists of a spin-wave state where one qubit per ensemble is flipped
\begin{align}
| \Psi_{\text{MF}}^{(1)} \rangle = \sum_{k=1}^M \psi_k | 1, \theta_k \rangle_k \prod_{i\ne k} | 0, \theta_i \rangle_i  , 
\label{firstexcited}
\end{align}
where we have defined 
\begin{align}
| 1, \theta \rangle & = \tilde{S}^x | 0, \theta \rangle_i .
\end{align}
and $ \tilde{S}^x = - \cos \theta S^x + \sin \theta S^z $ which creates a spin-flip in the $ \tilde{S}^z $-basis.  To apply the mean-field ansatz (\ref{firstexcited}) we diagonalize an effective Hamiltonian in the $ \psi_k $ coefficients and take the lowest energy state (see Appendix B).    We note that the mean-field theory is only expected to work in the regime with $ N \ge N_{\mathrm{c}} $, since the first excited state is taken to be of the form (\ref{firstexcited}), which has a spin configuration that is one spin-flip away from the ground state. 

The results are shown in Fig. \ref{fig2}(b), where the mean-field estimates (dashed lines) are compared to the exact results.  We see that excellent agreement in the energies of the states is obtained for all values of the adiabatic parameter $ \lambda $.  In Fig. \ref{fig2}(c) we plot the exact gap energy for various $ N $ in comparison to the mean-field calculation.   Figure \ref{fig2}(d) shows the convergence of the energies towards the mean-field results with $ N $ at various intermediate values of $ \lambda $. The mean-field results correspond to the limit $ N \rightarrow \infty $, and the exactly calculated gaps for various $ N $ rapidly approach the mean-field result. 

The results of Fig. \ref{fig2}(a)\--(d) were for a particular problem instance.  What is more meaningful is to study the performance of the scheme for a variety of different problem instances so that the overall behavior can be assessed. We find that the behavior is rather different depending upon whether $ N < N_{\mathrm{c}} $ or $ N \ge N_{\mathrm{c}} $, due to the different nature of the first excited state.  We study the two regimes separately by choosing problem instances where $ N_{\mathrm{c}} $ occurs relatively early ($N_{\mathrm{c}} = 3 $, Fig. \ref{fig2}(e)) or late ($N_{\mathrm{c}} > 7 $, Fig. \ref{fig2}(f)). In Fig. \ref{fig2}(e) we show the average, best, and worst scaling of the minimum gap for problems with $N_{\mathrm{c}}= 3$, such that most of the $N$-dependence is in the $ N \ge N_{\mathrm{c}} $ regime.  The best and worst scalings are defined as the largest and smallest difference in the gap comparing the qubit and $ N =7 $, the largest ensemble size calculated.  We find that the minimum gap increases with $ N $ on average.  Combined with the logically equivalent buffer states in the region of the ground state,  we expect that the AQC performance should improve for these cases.  

For the cases with $ N_{\mathrm{c}} > 7 $ (the $ N < N_{\mathrm{c}} $ regime), we see more mixed results (Fig. \ref{fig2}(f)). The average scaling tends to still improve with $ N $, but there are some cases where the minimum gap becomes significantly worse with $ N $.  In such cases we expect that the AQC performs poorly.  We note that the small values of $ N $ considered here are due to limitations in our numerical simulations. We thus expect that for realistic ensemble sizes would satisfy $ N \ge N_{\mathrm{c}} $, where the scaling is more favorable.  

This may, at first glance, seem to be a counter-intuitive result, since one might expect that with larger $ N $ the system should behave more classically.   However, it can be seen that in both (\ref{hzhamens}) and (\ref{hxhamens}) the magnitude of the elementary excitation does not diminish as $ N $ grow, since it is always a discrete Hamiltonian.  Thus, the gap does not diminish even for $ N \rightarrow \infty $, and AQC can be performed with macroscopically sized ensembles.

\begin{figure}
\begin{center}
\includegraphics[width=1\linewidth]{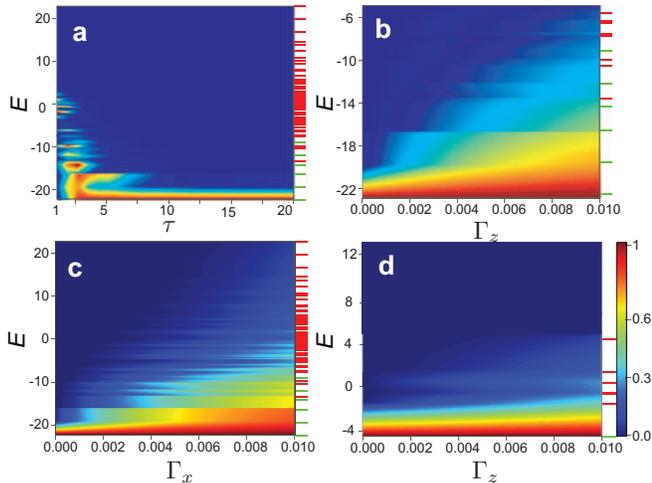} 
\end{center}
\caption{Occupation probability of the final state after time-evolving the adiabatic quantum computation. The occupation probability is plotted as a function of the (a) sweep time $ \tau $ with $ N = 5 $ and $\Gamma=0$; (b) $S^z$ decoherence rate $ \Gamma_z $ with $ N = 5 $; (c) $S^x$ decoherence rate $ \Gamma_x $ with $ N = 5 $; (d) $S^z$ decoherence rate $ \Gamma_z $ with $ N = 1 $. The same parameters as Fig. \ref{fig2}(b) are used $M=3 $ $J_{12}=-0.5, J_{13}=0, J_{23}=-1, K_1=0.5, K_2=0, K_3=1 $, and $\tau=100$. The probability distribution is normalized such that the maximal probability takes a value of 1 for each $ \tau $ or $ \Gamma_{x,z} $, for the clarity of the figure. For each plot, the spectrum of energy levels is labeled to the right of the plot.  Green levels indicate logically equivalent states to the ground state, red levels indicate logical error states.  \label{fig3}}
\end{figure}

\subsection*{Performance with adiabatic evolution}

We now directly time-evolve the AQC Hamiltonian and demonstrate its performance. We use a linear annealing schedule $ \lambda(t) = t/\tau $ and examine the final occupation probability of the states at time $ t = \tau $.  First, examining the case without decoherence, we vary the sweep time $ \tau $ for the same problem instance as shown in Fig. \ref{fig2}(b).  In Fig. \ref{fig3}(a), we see that as expected for sufficiently long $ \tau $, most of the population is concentrated in the ground state.  Diabatic excitations are seen for smaller $ \tau $ depleting the population in the ground state.  Here we also indicate the distribution of the energy levels that are logically equivalent to the ground state (green levels) and error states which have a different configuration (red levels) according to the majority vote encoding (\ref{equivalent}).  We see that the diabatic excitations tend to distribute the probability with a tendency to excite the lower energy states.  Since the logically equivalent states are also in the low-energy range, this shows that an effective buffer is provided by the encoding, protecting the computation as long as the diabatic excitations are within the logically equivalent states.  

Addition of decoherence has a qualitatively similar effect to diabatic excitations, as can be seen from Figs. \ref{fig3}(b)-(d). We numerically evolve a master equation in the presence of Markovian Lindblad dephasing in the $S^z $ and $S^x $ basis and obtain the final probability distributions for various decoherence rates (see Methods). 
Both the $S^z $- and $S^x $-dephasing is found to have a similar effect, with energy levels in the lower region being populated for stronger decoherence rates (Figs. \ref{fig3}(b)(c)).   Again, due to the fact that the logically equivalent states are dominated towards the lower end of the energy spectrum, this shows that the effect of the decoherence will also be to initially excite the logically equivalent states.   

It is interesting to compare the distribution to the unencoded bare AQC Hamiltonian $N = 1 $ (Fig. \ref{fig3}(d)) to the $ N = 5 $ encoded case (Fig. \ref{fig3}(b)), plotted on the same energy scale.  We see that for small decoherence rates, the effect of the dephasing is similar in terms of the width in the energy of the probability distribution. The benefit of the encoding scheme in the low decoherence region is evident from the fact that the first logical error state has a much higher energy in the $ N = 5 $ case rather than the $ N = 1 $ case (9 and 3 units above the ground state respectively). Summing the probability of all the states that are logically equivalent to the ground state (green levels), the ensemble case has a higher success probability. However, for larger dephasing rates, the distribution for the ensemble case becomes broader, and the error suppression is less effective. 
We therefore expect that as long as the decoherence rate is below a threshold, the overall logical errors can be effectively suppressed. 
We note that while we illustrate our results with a single problem instance, we have examined a variety of cases and seen consistently similar results for other problem instances. 

We now examine the dependence of the logical errors at the end of the AQC with the ensemble size $ N $.  In this case, it is illustrative to examine another problem instance, corresponding to the ferromagnetic Hamiltonian $ J_{ij} = -1 ( 1- \delta_{ij}) $ with a bias field  $ K_i = K $. For $ N =1  $ with $ K > 0 $, the ground state is $ \sigma_i = - 1 $ and the first excited state is $ \sigma_i = + 1 $ for all $ i $. The energy landscape corresponds to one global minimum and one local minimum separated by a potential barrier consisting of all the remaining states.   For the ensemble case and in the regime $ N \ge N_{\mathrm{c}} $, the ground state has the same logical configuration $ S^z_i = - N $ but the first excited state is a single spin-flip of the configuration $ S^z_k = - N+2, S^z_i = - N $ $\forall i \ne k $ (see Appendix C). The error probability is defined as $1-(\text{success probability})$, where the success probability is defined to be the total probability of all states that are logically equivalent to the ground state.

\begin{figure*}
\begin{center}
\includegraphics[width=0.9\linewidth]{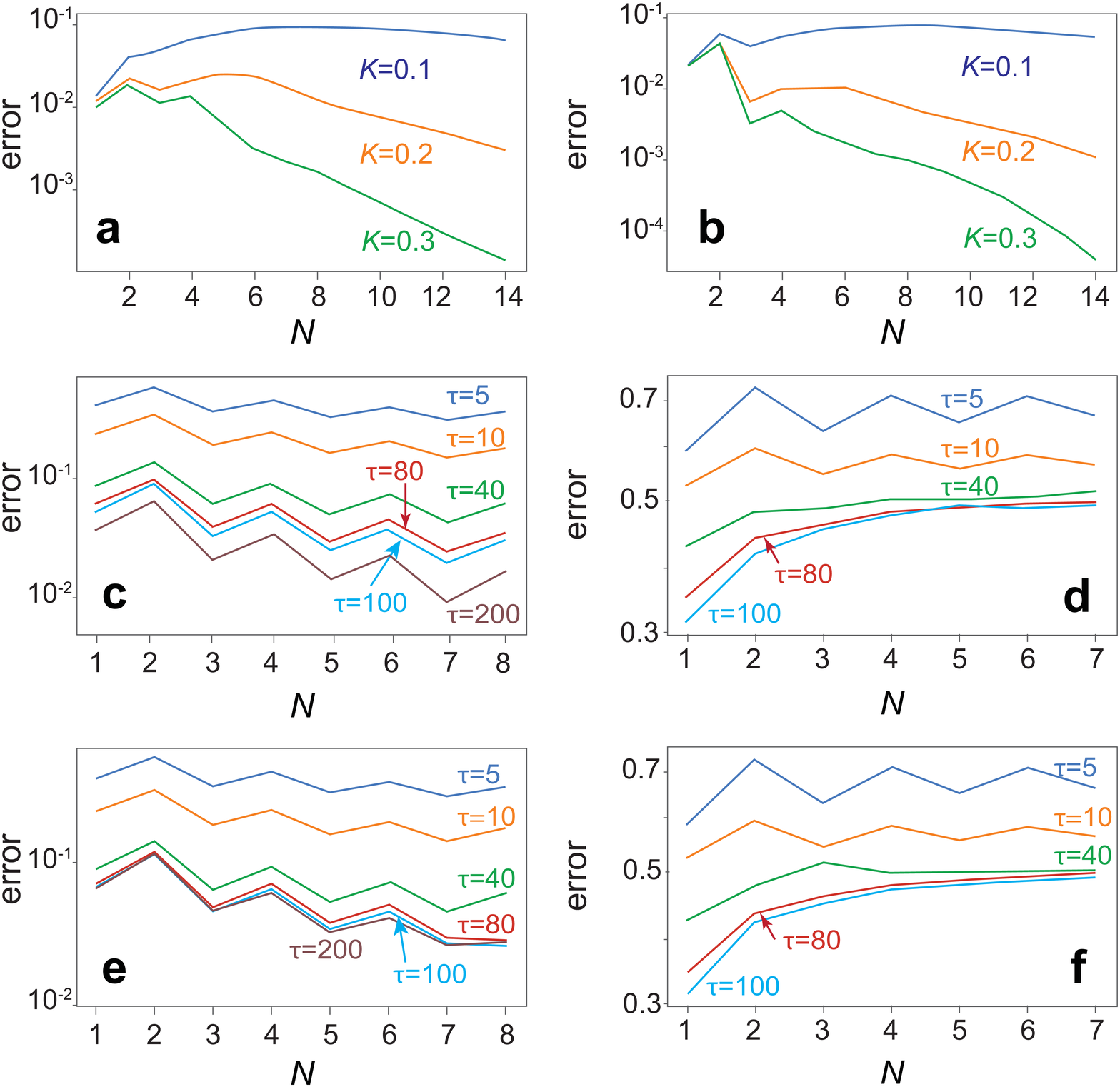} 
\end{center}
\caption{Error probabilities in the time-evolved adiabatic quantum computation.  (a)(b)  Error versus $ N $ for the ferromagnetic model with $ M = 3 $, $ J_{ij} = -1 ( 1- \delta_{ij}) $ and $ K_i = K $, and $ \tau = 100 $.  The critical ensemble size for each parameter is:  $ N_{\mathrm{c}} (K=0.1) = 14 $, $ N_{\mathrm{c}} (K=0.2) = 8 $, $ N_{\mathrm{c}} (K=0.3) = 5 $.  Dephasing in the basis (a) $ S^z $-basis with $ \Gamma^z = 10^{-4}$ and (b) $ S^x $-basis with $ \Gamma^x = 10^{-4}$ are calculated.  (c)(d) Averaged error versus $N$ for various $\tau$ and $ M=3 $  for (c)  60 problem instances with $ N_{\mathrm{c}} = 3 $;  (d) 60 problem instances with $ N_{\mathrm{c}} > 7 $.   Averaged error versus $N$ for various $\tau$ and $ M=3 $ including dephasing of rate $\Gamma=10^{-4}$ for the same problem instances with (e) $N_{\mathrm{c}}=3 $;  (f) $ N_{\mathrm{c}} > 7 $. \label{fig4}}
\end{figure*}

Figure \ref{fig4}(a)(b) show the error probabilities for $ S^z $ and $S^x $ dephasing, respectively for the ferromagnetic instance with different values of $K$. We observe that the error probabilities increase initially, but start to strongly decrease as $ N $ is increased further.  While lower error probabilities are expected from a larger gap with larger $ N $ as observed in Fig. \ref{fig2}(e), the decrease in error is much more than would be expected from this. The suppression of the error with increasing $ N $ attributed to the fact that above $N_{\mathrm{c}}$,  the first excited state becomes a logically equivalent state, which is one of the characteristics of the $ N \ge N_{\mathrm{c}} $ regime (Fig. \ref{fig1}(a)). As the effect of the dephasing is to excite low-lying energy states, the excitation of the first excited state no longer becomes a logical error, suppressing the total error. Such an improvement is consistent with analysis in Ref. \cite{vinci2018scalable} where the encoding was shown to give effectively give lower temperatures, and Ref. \cite{PhysRevA.99.062307} where phase transitions in the model were shown to be weakened.

The phenomenology of an error increase in the region for small $ N $  but decreasing error for large $N $ is also observed in other problem instances. We randomly generate problem instances and calculate the error probability of the ensemble encoding as a function of various ensemble sizes $ N $ and sweep times $ \tau $ in the absence and presence of decoherence. For each generated problem instance, we calculate the critical $ N_{\mathrm{c}} $, and again group 
the instances according to whether $ N_{\mathrm{c}} $ occurs relatively early (specifically $ N_{\mathrm{c}} = 3 $) or late ($ N_{\mathrm{c}} > 7 $).  Examining each case separately, we can study the performance of the ensemble scheme in the $ N \ge N_{\mathrm{c}} $ or $ N< N_{\mathrm{c}} $ regime, respectively. Fig. \ref{fig4}(c)-(f) shows the error probability of finding the state of the system at the end of adiabatic evolution, averaged over a set of fixed 60 randomly chosen problem instances with $ N_{\mathrm{c}} $ in each range.

First examining  problems with $N_{\mathrm{c}} =3$ (the $ N \ge N_{\mathrm{c}} $ regime) without decoherence, we observe that for sufficiently long $ \tau $ the error probability strongly decreases with $ N $ (Fig. \ref{fig4}(c)).  We again attribute this to the presence of logically equivalent excited states in the vicinity of the ground state. Although we only simulated relatively small system sizes due to numerical limitations, we expect that 
for larger $ N $ the trend will continue towards lower errors as the gap energy approaches the mean-field value corresponding to $ N \rightarrow \infty $.  
For short sweep times $ \tau $, the errors also improve with $ N $, although with a smaller gradient.  We attribute this behavior due to the sweep times being in a diabatic regime, such that the system is not maintained in the vicinity of the ground state, which involves high energy excitations.  We note that there did exist rare examples where the error probability scaled badly with $ N $, due to the particular structure of the energy spectrum.  However, the occurrence of these poorly scaling examples were so rare that they did not impact the average to a significant extent. 

For problems with $N_{\mathrm{c}}>7$ (the $ N < N_{\mathrm{c}} $ regime) and no decoherence, the error tends to increase with $ N $ (Fig. \ref{fig4}(d)), despite the fact that the average minimum gap increases with $ N $, as seen in Fig. \ref{fig2}(f). We have examined the individual cases and confirmed that for particular cases where the gap increases with $ N $, the error decreases with $ N $ as expected.  The reason that the average error increases is that the cases with poor gap scaling in Fig. \ref{fig2}(f) tend to have nearly zero success probability, which reduces the average, considerably.  Thus, the results for problems in the $N < N_{\mathrm{c}}$ regime are mixed and depend very much upon the particular problem instance of whether the gap increases or decreases.  We do, however, note that these problem instances themselves are rather rare, as seen in Fig. \ref{fig1}(c), in comparison to the more common $ N > N_{\mathrm{c}} $ case.  When averaged over all random problem instances, errors tend to decrease with $ N $.  

Calculations adding $ S^z $-dephasing to the AQC scheme is shown in Fig. \ref{fig4}(e)(f). We find generally the same behavior of the error with $ N $ when decoherence is introduced, but with a higher error probability overall, as expected. For the $N_{\mathrm{c}} =3$ case shown in Fig. \ref{fig4}(e), we see that there is a similar improvement of the error with $ N $ as the zero decoherence case.  The new feature here is that there is an optimum sweep time beyond which the error probability starts increasing again.  This can be simply explained by noting that the  AQC must be performed within the decoherence time available to the computation.  Beyond the optimal time, the performance starts to degrade, therefore there is a trade-off between maintaining adiabaticity and working within the decoherence time \cite{PhysRevA.65.012322,keck2017dissipation}.   For the $N_{\mathrm{c}} > 7 $ cases in Fig. \ref{fig4}(f), we again see the error increase with $ N $, which is attributed again to particular instances where the minimum gap decreases with $ N $.  

  
 We note that in Figs. \ref{fig3} and \ref{fig4} we have used collective dephasing with respect to the $ S^z $ and $ S^x $ operators, which is one of the decoherence channels for the atomic ensemble implementation \cite{byrnes2015macroscopic,PhysRevA.88.023609,riedel2010atom,RevModPhys.90.035005}.  For other implementations individual qubit dephasing may be a more relevant decoherence model. We generally expect similar results with individual qubit dephasing, since elementary perturbations due to decoherence produce similar transitions in the energy spectrum of the Hamiltonian.  Our calculations based on the ferromagnetic model confirm that qualitatively similar results are obtained with an individual qubit dephasing model.  The primary difference in this case is the addition of non-symmetric states, which can also act as buffer states to protect the ground state (see Appendix D). 

We finally comment on the presence of entanglement during the AQC sweep. 
The mean-field wavefunction as given in (\ref{ground}) takes the form of a product state of spin coherent states on each ensemble. This may suggest that there is no entanglement between the ensembles during the adiabatic evolution.  In fact, entanglement is typically present during the evolution, due to the $ S^z_i S^z_j $ interaction in the AQC Hamiltonian.  The mean-field ground state is merely an approximation to the true ground state, which in fact typically contains entanglement. We have explicitly calculated entanglement for small ensemble sizes (see Appendix E).  The presence of entanglement is consistent with past works studying the robustness of entanglement in the presence of decoherence \cite{PhysRevA.88.023609,pyrkov2013entanglement,RevModPhys.82.1041,julsgaard2001experimental, sherson2006quantum,kunkel2017,lange2017,fadel2017}.  The factor of $ 1/N $ multiplying the $ S^z_i S^z_j $ terms makes the type of entangled state of a robust type as discussed in Ref. \cite{PhysRevA.88.023609}.  We therefore, expect that the entanglement should survive for macroscopic ensembles within the decoherence window.  This can be contrasted to other ensemble-based approaches such as in liquid-state NMR \cite{PhysRevLett.88.167901,PhysRevLett.83.1054}, where the entanglement collapses to zero.

 

\subsection*{Experimental implementation}

We briefly describe a potential experimental implementation which realizes the Hamiltonians (\ref{hzhamens}) and (\ref{hxhamens}). Neutral atom ensembles, consisting of either thermal atomic ensembles or cold atoms, are a strong candidate for realizing a single spin ensemble $ S_i^{x,z} $  \cite{RevModPhys.90.035005,RevModPhys.82.1041,julsgaard2001experimental,sherson2006quantum,abdelrahman2014coherent}.  The individual qubits of spin $ \sigma_{i,n}^{x,z} $ within the ensemble are realized by the internal hyperfine ground states of the atoms.  For example, for $ ^{87} \text{Rb} $ the ground states $ F=1,  m_F = -1 $ and $ F=2, m_F = 1 $ are clock states such as their energy separation is insensitive to magnetic field fluctuations \cite{RevModPhys.90.035005}.  Thermal atomic ensembles are trapped in paraffin-coated glass cells which prolong the coherence time of the internal states \cite{RevModPhys.82.1041}.  In this case, each glass cell acts as one of the ensemble spins $ S_i^{x,z} $, such that $ M $ glass cells are prepared.  Such a multi-ensemble system was realized in Ref. \cite{pu2017experimental} where 225 locally addressable atomic ensembles were created, as well as entanglement generation between 25 ensembles \cite{pu2018experimental}.  Another approach is to have a multi-trap atom chip system, such as that proposed in Ref. \cite{abdelrahman2014coherent,byrnes2015macroscopic}.  Atom chips are a flexible platform for producing magnetic traps, where atoms can be cooled to quantum degeneracy.  One advantage of cooling is that long coherence times can be achieved (up to 1 minute in Ref. \cite{deutsch2010spin}), for atomic gases just above the critical temperature for Bose-Einstein condensation.   

To realize Hamiltonians (\ref{hzhamens}) and (\ref{hxhamens}), we require both ensemble-ensemble effective interactions and single ensemble control. This can be produced by using optically mediated methods, where off-resonant lasers produce an effective $ S_i^z S^z_j $ type interaction \cite{pyrkov2013entanglement,hussain2014geometric,pettersson2017light}. An alternative for cold atom systems is to produce the interactions by taking advantage of the non-linear interactions between the atoms using state-dependent forces \cite{treutlein2006microwave}.  Optically mediated methods are flexible as they are able to produce remote entanglement, whereas the interaction methods require bringing the ensembles into close proximity. Adjusting the interaction parameters allows one to realize a controllable $ J_{ij} $ matrix between the ensembles. The  Hamiltonian (\ref{hxhamens}) can be achieved by microwave/radio frequency transitions.  In the case of $ ^{87} \text{Rb} $, this is realized by a two-photon transition between the clock states \cite{riedel2010atom}.  Unlike optical frequencies which can be focused on a single ensemble, microwave/radio frequency transitions are less easily directed and are applied on all ensembles in the same vicinity, which is sufficient for the $ S^x_i $ Hamiltonian.  Finally, the $ S^z_i $ terms (\ref{hzhamens}) could be realized by a state-dependent potential \cite{bohi2009coherent} or an ac Stark shift to optically shift the energy levels \cite{abdelrahman2014coherent}.  We note that other physical systems could potentially implement the Hamiltonians  (\ref{hzhamens}) and (\ref{hxhamens}), for example, ensembles of NV-centers \cite{bar2013solid,pham2011magnetic,stanwix2010coherence} are another possibility.  

One of the attractive features of the AQC Hamiltonian (\ref{hzhamens}) and (\ref{hxhamens}) is that it is written completely in terms of the total spin of the ensembles of qubits.  This means that only {\it global} control  of the ensembles --- involving the collective spin operators $ S^{x,z}_i $ --- rather than the control of {\it individual} qubits $ \sigma^{x,z}_{i,n} $ is necessary.  Alleviating the need for individual qubit control in realizing an error protected quantum computer is a significant simplification, since most quantum error-correction schemes require rather sophisticated degrees of quantum control \cite{RevModPhys.87.307,devitt2013quantum}.  The resources required to scale up the degree $ N $ of the repetition code would be considerably less in an approach involving only collective spin control. For example, in atomic ensembles on atom chips, one typically deals with atom clouds of the order $ N \sim 10^3 $ \cite{riedel2010atom,fadel2017}, and in paraffin coated glass cells the number of atoms is $ N \sim 10^{12} $ \cite{julsgaard2001experimental}, which are all manipulated in parallel within the experiment.   
In systems where each of the copies must be implemented with individual qubit control, the resources are consequently greater.  In comparison, past demonstrations on the D-wave machines have been in the region of $ N \sim 10 $ \cite{pudenz2014error,venturelli2015quantum,vinci2016nested}.  While related approaches have been considered in the context of all-to-all encodings such as nested QAC \cite{venturelli2015quantum,vinci2016nested,vinci2018scalable}, the chimera graph hardware that was used necessitates a further minor embedding step which is not required in our case.

An important issue when using macroscopic ensembles is sensitivity to decoherence. One might naively expect that atomic ensembles containing up to $ N \sim 10^{12} $ atoms are very sensitive to decoherence. In fact, the sensitivity of the quantum state to decoherence is a highly state-dependent process \cite{julsgaard2001experimental}.  For example, cat states are extremely sensitive to decoherence, but spin coherent states are more robust \cite{PhysRevA.88.023609,byrnes2020quantum}.  Hence the important consideration is the type of states that are generated, and one should be careful that they are robust in the presence of decoherence.  It was shown in Ref. \cite{PhysRevA.88.023609} that the $ S_i^z S^z_j $ interaction produces states which are robust in the presence of decoherence as long as the interaction timescale is of order $ \sim 1/N $.  In this respect the factor of $ 1/N $ in front of the two-ensemble interaction (\ref{hzhamens}) is beneficial as it implies highly decoherence-sensitive states are not generated.  Another indication of this is that in the $ N \ge N_{\mathrm{c}} $ regime, the mean-field state (\ref{ground}) is a tensor product of spin coherent states, which are known to be robust in the presence of decoherence.  
This explains the good performance of our proposed scheme even in the presence of dephasing, as shown in Fig. \ref{fig4}.  

Further details of the implementation with neutral atom ensembles can be found in Refs. \cite{byrnes2012macroscopic,abdelrahman2014coherent,byrnes2015macroscopic} and Appendix F.

\section*{DISCUSSION}

In this study, we investigated a formulation of AQC where qubit ensembles are used instead of qubits, and the ensemble Hamiltonians (\ref{hzhamens}) and (\ref{hxhamens}) are adiabatically evolved. We have found that finding the ground state of (\ref{hzhamens}) is an equivalent problem to the original qubit problem Hamiltonian (\ref{hzham}).  
The main difference of the ensemble and qubit problem Hamiltonians is that the ensemble version introduces many logically equivalent states as defined in (\ref{equivalent}) with similar energies to the ground state.  The introduction of these states is beneficial for AQC since occupation of these  states do not cause a logical error, and provide a buffer against diabatic excitation.  We found that there are two important regimes with respect to $N $, depending on the character of the first excited state, summarized in Fig. \ref{fig1}(a). In the regime with $ N \ge N_{\mathrm{c}} $, we find that the minimum gap energy increases, and the ground state is logically protected, leading to a reduced error probability in the AQC. In the regime with $ N < N_{\mathrm{c}} $, we obtain mixed results, where despite the average  minimum gap increasing, the AQC scales on average poorly.  This was due to the particularly poor performance of the cases where the gap decreases, and can be attributed to the lack of logical protection of the ground state.  For large ensemble sizes such as that realized with atomic ensembles, all but a minority of problems should satisfy $ N \ge N_{\mathrm{c}} $, where the ground state is logically protected. 

We thus find that AQC with ensembles should perform well in a great majority of cases for large $ N $.  
One may find it surprising that it is possible to perform AQC at all with ensembles of qubits, even in the limit of $N \rightarrow \infty $.  The first key point that allows for the ensemble version to still work is that the discrete nature of the Hamiltonian is preserved under  (\ref{hzhamens}) and (\ref{hxhamens}).  Thus, although the energy of the full space can be viewed as being quasi-continuous as shown in Fig. \ref{fig1}(a), this is only because the space is being viewed in rescaled variables $ x_i = S_i^z/N $. Physically, the magnitude of the spins are also growing with $ N $, which preserves the energy gap.  From a resource point of view, one may argue that many more physical qubits are being used.  However, we take the point of view that the relevant resource is the complexity of the experiment control when dealing with an ensemble as compared to a  single qubit. 
For many implementations the effort required for controlling an ensemble is no more than that of a single qubit.  For instance, if performing a single qubit operation on an atom is performed by a laser pulse, then the equivalent ensemble operation is to apply the pulse on the whole ensemble.  This is typically not an operation that costs  $ N $ times more since one can illuminate the whole ensemble with the same laser, i.e. it is parallelizable.    Thus, as long as the operations for the qubit operators $ \sigma^{x,y,z}_i $ can be performed with a similar experimental overhead to ensemble operators $ S^{x,y,z}_i $, then implementing the ensemble and qubit version of the AQC Hamiltonians should be of comparable complexity.  

One may also be concerned that the use of ensembles may be problematic since they could be extremely sensitive to decoherence, owing to their macroscopic nature.  The sensitivity of qubit ensemble states has already been investigated in numerous works, see for example Refs. \cite{PhysRevA.88.023609,byrnes2012macroscopic,semenenko2016implementing,pyrkov2014quantum}. The main point here is that the fragility of the quantum states is state-dependent:  while Schrodinger cat states are extremely sensitive, spin coherent states are generally quite robust.  This is what has allowed the experimental realization of macroscopic quantum states, such as those performed by Polzik and co-workers \cite{RevModPhys.82.1041,julsgaard2001experimental, sherson2006quantum}.  The form of the mean-field ground and excited states suggest that the ensemble version of AQC can also be robust for the same reasons.  The ground state (\ref{ground}) is nothing but a set of spin coherent states, and (\ref{firstexcited}) is a spin-wave excitation on the ground state.  Spin-wave states are also relatively robust and have been already demonstrated experimentally \cite{bao2012quantum}. Therefore, as long as the ensemble size is such that $ N \ge N_{\mathrm{c}} $, we believe that it is reasonable to expect that the scheme works even in the presence of decoherence.  If it is the case that $ N < N_{\mathrm{c}} $, it is less clear what the decoherence properties are since the nature of the state is not yet understood.  We nevertheless observe that in some cases the minimum gap can increase, making the ensemble framework viable.  While we have not been able to exactly characterize the cases that are most susceptible, we also have not seen any correlation with classically hard instances of combinatorial problems (see Appendix D).  Considering that these are a small fraction of the full problem set for large $ N $, we find that in most cases the ensemble framework successfully performs error-suppression via the duplication of the quantum information.  This is consistent with other approaches using repetition codes with AQC \cite{pudenz2014error,venturelli2015quantum,vinci2016nested,PhysRevLett.116.220501,PhysRevA.95.022308,mishra2016performance,vinci2015quantum,pearson2019analog,vinci2018scalable,PhysRevA.99.062307}.  

Another direction that could be further investigated is the use of energy penalty terms, which have been shown to be beneficial in several works \cite{venturelli2015quantum,vinci2016nested,vinci2018scalable,PhysRevA.99.062307,pudenz2014error}.  This would involve the addition of terms to the Hamiltonian of the form $ (S^z_i)^2 $ to induce a ferromagnetic interaction within the qubits in the ensembles.  This is expected to further improve the performance, where there is an optimal strength of the interaction.  Producing such an interaction in the case of atomic ensembles has been experimentally and theoretically investigated \cite{riedel2010atom,pyrkov2013entanglement}, and is compatible with the general framework of our approach since it is based on collective operations of the ensemble.  We will leave these topics for further investigation as future work.

\section*{Methods}
\subsection*{Numerical simulation}

To examine the performance of the ensemble version of AQC, we performed both a pure state evolution and mixed state evolution of Hamiltonian (\ref{aqcham}). We use a linear annealing schedule $ \lambda(t) = t/\tau $ and examine the final occupation probability of the states at time $ t = \tau $.  For the  case that we include decoherence, we consider Markovian $S^z$- and $S^x$-dephasing.  This is particularly relevant for an implementation with atomic ensembles, where the coupling to the ensemble spins occur in a collective manner \cite{byrnes2015macroscopic,PhysRevA.88.023609,riedel2010atom,RevModPhys.90.035005}.  We use the master equation \cite{navarrete2015open}
\begin{align}
\frac{d \rho}{dt}= & i [\rho,H]
-\frac{\Gamma_z}{2}\sum_{n=1}^{M} [\rho (S^z_n)^{2}- 2 S^{z}_n \rho S^{z}_n+ (S^{z}_n)^{2}\rho ]\nonumber \\
& -\frac{\Gamma_x}{2}\sum_{n=1}^{M} [\rho (S^x_n)^{2}- 2 S^{x}_n \rho S^{x}_n+ (S^{x}_n)^{2}\rho],
\label{collectivedephase}
\end{align}
where $\Gamma _{z,x}$  are dephasing rates and $ H $ is the Hamiltonian (\ref{aqcham}). Starting from the eigenstate of the initial Hamiltonian we solve the master equation numerically for the combined adiabatic and dephasing evolution. The performance of the AQC is then evaluated through the probability of finding the state of the system in the ground state at the end of the adiabatic evolution. Qutip was used for the simulations \cite{johansson2012qutip}.

%
\section*{Acknowledgments}
This work is supported by the National Natural Science Foundation of China (62071301); State Council of the People's Republic of China (D1210036A); NSFC Research Fund for International Young Scientists (11850410426); NYU-ECNU Institute of Physics at NYU Shanghai; the Science and Technology Commission of Shanghai Municipality (19XD1423000); the China Science and Technology Exchange Center (NGA-16-001). J. P. D. would like to acknowledge support from the US Air Force Office of Scientific Research, the Army Research Office, the Defense Advanced Funding Agency, the National Science Foundation, and the Northrop-Grumman Corporation.  A. P. acknowledges the RFBR-NSFC collaborative program (Grant No. 18-57-53007).


\appendix

\section{Equivalence of ground states between qubit and ensemble $ H_Z $}
\label{sec:equivall}

\subsection{Equivalence of states at corners of hypercube}
\label{sec:equiv}

Dividing (4) in the main text by $ N $, we obtain
\begin{align}
E(x_1, \dots, x_M) &  = \frac{H_Z}{N} \nonumber \\
& = \sum_{i,j=1}^M J_{ij} x_i x_j + \sum_{i=1}^M K_i x_i
\label{hzovern}
\end{align}
where we have defined
\begin{align}
x_i = \frac{S_i^z}{N}  .
\end{align}
We will work with rescaled energies $ E $ that divide the Hamiltonian (4) in the main text by $ N $. The eigenvalues of $ S^z $ take the values $ \{-N, -N+2, \dots, N \} $.  Consequently the eigenvalues of $ x_i $ take the values $ \{-1, -1+2/N, \dots, 1 \} $.  Comparing (\ref{hzovern}) to (1) in the main text, since the eigenvalues of $ \sigma^z_i $ are $ \{ -1, 1 \} $, for $ x_i $ values taking $ \{ -1, 1 \} $ (i.e. the hypercube corners), the same energy is obtained up to a constant factor of $ N $. The spin configurations are thus equivalent in the sense of (7).

\subsection{Variation of the energy along the edge of the hypercube}
\label{sec:edge}

Let us now see the variation of the energy $ E $ where we start from the state that is equivalent to the qubit ground state configuration $ x_i = \sigma_i $. The qubit ground state energy is
\begin{align}
E_0 &  = \sum_{i,j=1}^M J_{ij}  \sigma_i   \sigma_j + \sum_{i=1}^M K_i  \sigma_i .
\label{groundstateen}
\end{align}
Now parametrize the deviation from the ground state hypercube corner using
\begin{align}
x_i = \sigma_i (1- 2 \epsilon_i)
\label{xparam}
\end{align}
where $ \epsilon_i \in [0,1] $.  The energy for an arbitrary deviation from the ground state hypercube corner is
\begin{align}
E(\epsilon_1, \dots, \epsilon_M)  = & \sum_{i,j=1}^M J_{ij} \sigma_i \sigma_j (1- 2 \epsilon_i)(1- 2 \epsilon_j) \nonumber \\
& + \sum_{i=1}^M K_i \sigma_i (1- 2 \epsilon_i).
\end{align}
The energy variation changing just one of the spins $ x_k $ gives a rate of change in the energy 
\begin{align}
\frac{\partial E}{\partial \epsilon_k} = -2 \sigma_k \left( 2 \sum_i J_{ik} x_i +  K_k \right)  .
\label{edgevariation}
\end{align}
This is a constant with respect to $ \epsilon_k $, hence we can observe that the energy variation when changing only one of the spins is always linear. A special case of this when moving along one of the hypercube edges where $ x_i = \pm 1 $.  The variation in energy when moving alone a hypercube edge is always linear. 

The ground state configuration corresponds to $ \epsilon_i = 0 $. Starting from the ground state, the energy variation changing one of the spins is thus
\begin{align}
E(0, \dots,0, \epsilon_k,0, \dots, 0) = E_0 + \frac{\partial E}{\partial \epsilon_k} \Bigg|_{\bm{\epsilon} =0}  \epsilon_k .
\end{align}
This corresponds to moving along one of the edges of the hypercube, starting from the equivalent ground state configuration.

Varying $ \epsilon_k $ from 0 to 1 and
keeping all the other $  \epsilon_i = 0 $ corresponds to flipping one of the spins from $ x_k \rightarrow - x_k $.  This is another hypercube corner, which has an equivalent spin configuration according to the result in Sec. \ref{sec:equiv}.  Since hypercube corners have the same energy $ E $ as the original qubit problem, this is guaranteed to have a higher energy since the original state $ \epsilon_i = 0 $ is by definition the ground state.  Combining this with the fact that the energy variation along a hypercube edge is linear, we conclude that
\begin{align}
\frac{\partial E}{\partial \epsilon_k} \Bigg|_{\bm{\epsilon} =0} = -2 \sigma_k \left( 2 \sum_i J_{ik} \sigma_i +  K_k \right)  \ge 0 .
\label{gradientatgs}
\end{align}

The gradient in the vicinity of the ground state is then
\begin{align}
\bm{\nabla} E \big|_{\bm{\epsilon} =0} = \left( \frac{\partial E}{\partial \epsilon_1}, \frac{\partial E}{\partial \epsilon_2}, \dots, \frac{\partial E}{\partial \epsilon_M}  \right) \Bigg|_{\bm{\epsilon} =0} .
\end{align}
The energy variation in an arbitrary direction starting from the ground state corner is therefore given by 
\begin{align}
E(\epsilon_1, \epsilon_2, \dots, \epsilon_M) \approx E_0 + \bm{\nabla} E \big|_{\bm{\epsilon} =0}  \cdot (\epsilon_1, \epsilon_2, \dots, \epsilon_M) .
\end{align}
From (\ref{gradientatgs}), since each of the derivatives are positive, we can conclude that the gradient of the energy in an arbitrary direction in the vicinity of the ground state will always be positive.

\subsection{Variation of energy from the ground state to an arbitrary hypercube corner}

The previous section shows that the energy increases in an arbitrary direction starting from the ground state corner. We also showed that varying the energy along an hypercube edge changes the energy linearly.  We now examine the energy variation starting from the ground state hypercube corner to an arbitrary hypercube corner. 
A linear trajectory connecting the ground state corner to an arbitrary corner is defined by
\begin{align}
\epsilon_i = n_i \varepsilon
\label{inttraj}
\end{align}
where $ n_i \in \{0,1\} $ are integer parameters which determine the trajectory and $ \varepsilon \in [0,1] $ is the parameter which determines the position along the chosen trajectory.  

Substituting (\ref{inttraj}) and (\ref{xparam}) into (\ref{hzovern}) and subtracting off the ground state (\ref{groundstateen}), we obtain
\begin{align}
f_{n_1 \dots n_M}&  (\varepsilon)  = E(x_1, \dots, x_M)  - E_0 \nonumber \\
& = - 4\sum_{i,j} J_{ij} \sigma_i \sigma_j \left( n_i \varepsilon - n_i n_j \varepsilon^2 \right) - 2 \sum_i K_i \sigma_i n_i \varepsilon .
\label{ffunc}
\end{align}

We know that when $ \varepsilon = 1 $, this corresponds to another hypercube corner. From the result of Sec. \ref{sec:equiv}, this must have a higher energy since $ \varepsilon = 0 $ has by definition the same energy as the qubit ground state.   Furthermore, we observe from (\ref{ffunc}) that the variation is always a quadratic polynomial with respect to $ \varepsilon $.  

We now show that there is no other lower energy state along the line connecting the ground state to another hypercube corner. From the fact that the energy gradient is positive from (\ref{gradientatgs}), and the energy varies quadratically, there are only two possibilities for the type of curve that (\ref{ffunc}) follows with respect to  $ \varepsilon $.  First consider the case that the parabola in  $ \varepsilon $ is concave upwards, namely 
\begin{align}
\sum_{i,j} J_{ij} \sigma_i \sigma_j  n_i n_j \ge 0  .
\end{align}
Since the energy gradient at $ \varepsilon = 0 $ is positive, the turning point must occur for $ \varepsilon< 0 $ and we can deduce that the energy monotonically increases to the hypercube corner.  In the case that the parabola is concave downwards, namely
\begin{align}
\sum_{i,j} J_{ij} \sigma_i \sigma_j  n_i n_j < 0 ,
\end{align}
then for the same reasons, this means that the turning point occurs for $ \epsilon > 0 $.  Since the hypercube corner with $ \varepsilon =1  $ is at a higher energy than the ground state, the energy rises monotonically if the turning point is $ \varepsilon \ge 1 $.  If the turning point is $0 <  \epsilon < 1 $, the energy increases, then decreases and there is a maximum in the energy between the hypercube corners.  In all the cases the minimum is at the  ground state hypercube corner, i.e. $ \varepsilon =0  $ . 

Since $ f_{n_1 \dots n_M} (\varepsilon =0) = 0 $ by construction, the above result implies that
\begin{align}
f_{n_1 \dots n_M} (\varepsilon) \ge 0  .
\end{align}

\subsection{Variation of energy from the ground state in an arbitrary direction}

We have so far shown that energy of any of the points along the trajectory connecting the ground state hypercube corner to any other hypercube corner is higher or the same as the ground state energy $ E_0 $.  To show an arbitrary point in the hypercube also has a higher energy than $ E_0 $, we should connect the ground state along an arbitrary trajectory through the hypercube.  This can be parameterized by
\begin{align}
\epsilon_i = \alpha_i \varepsilon.
\label{inttrajalpha}
\end{align}
where in this case $ \alpha_i \in [0,1] $ are continuous parameters that determine the trajectory. In order to have $ \varepsilon \in [0,1] $ be the full range of the trajectory, we demand that at least one of the $ \alpha_i  $ be equal to unity, i.e. 
\begin{align}
\sup_i  \alpha_i  = 1.
\end{align}
The energy variation along this trajectory can be calculated in a similar way to (\ref{ffunc}), and is given by 
\begin{align}
F(\varepsilon) & = E(x_1, \dots, x_M)  - E_0 \nonumber \\
& = - 4\sum_{i}  \sum_{j\ne i} J_{ij} \sigma_i \sigma_j \left( \alpha_i \varepsilon - \alpha_i \alpha_j \varepsilon^2 \right) - 2 \sum_i K_i \sigma_i \alpha_i \varepsilon \nonumber \\
& = \sum_{i} \sum_{j\ne i}  D_{ij} ( \varepsilon )  \alpha_i \alpha_j + \sum_i C_i \alpha_i,
\label{ffuncalpha}
\end{align}
where we have defined
\begin{align}
D_{ij} ( \varepsilon ) & = 4  \varepsilon^2 J_{ij} \sigma_i \sigma_j,  \nonumber \\
C_{i} ( \varepsilon ) & = - 2 \varepsilon \left( K_i\sigma_i  + 2 \sum_{j \ne i } J_{ij} \sigma_i \sigma_j  \right),
\end{align}
and used the fact that $ J_{ii} = 0 $.

We would like to show that $ F(\varepsilon) \ge 0 $ for any choice of $ \alpha_i $, which would show that a state along an arbitrary trajectory starting from the ground state hypercube corner always has a higher or the same energy.  To achieve this, we construct the function $ F(\varepsilon) $  with a linear combination of the basis functions $ f_{n_1 \dots n_M} (\varepsilon) $. That is, we require coefficients such that 
\begin{align}
F(\varepsilon) = \sum_{n_1,\dots,n_M} w_{n_1 \dots n_M} f_{n_1 \dots n_M} (\varepsilon),
\label{expansionF}
\end{align}
where $ n_i \in \{0,1 \} $.  Since the basis functions are individually positive (\ref{ffunc}), it then follows that if 
\begin{align}
w_{n_1 \dots n_M}  \ge 0, 
\label{positivec}
\end{align}
then  $ F(\varepsilon) \ge 0 $. 

Substituting (\ref{ffunc}) and (\ref{ffuncalpha}) into (\ref{expansionF}), we obtain
\begin{align}
\sum_{n_1,\dots,n_M}&  w_{n_1 \dots n_M} f_{n_1 \dots n_M} (\varepsilon) \nonumber \\
& = \sum_{i} \sum_{j\ne i}  D_{ij} ( \varepsilon ) \sum_{n_1,\dots,n_M} w_{n_1 \dots n_M} n_i n_j \nonumber \\
& + \sum_{i} C_{i} ( \varepsilon ) \sum_{n_1,\dots,n_M} w_{n_1 \dots n_M} n_i .
\label{summedf}
\end{align}
Comparing (\ref{summedf}) and (\ref{ffuncalpha})  we observe that firstly the coefficients must satisfy for $ i \ne j $
\begin{align}
\sum_{n_1,\dots,n_M} w_{n_1 \dots n_M} n_i n_j = \alpha_i \alpha_j .
\label{eps2sum}
\end{align}

This is a set of $ M^2-M $ linear equations to be solved with $ 2^M $ variables.  There are more degrees of freedom than equations, hence will typically be many choices of $  w_{n_1 \dots n_M} $ that satisfy (\ref{eps2sum}). Here we note that in this sum, any $ n_i $ which has a single ``1'' and all others ``0''
\begin{align}
\sum_i n_i =1,
\label{deltani}
\end{align}
does not contribute due to the product $ n_i n_j $, noting that $ i \ne j $. 

 We also require that 
\begin{align}
\sum_{n_1,\dots,n_M} w_{n_1 \dots n_M} n_i  = \alpha_i ,
\label{epsconst}
\end{align}
which is another set of $ M $ equations to be satisfied. This set of equations can be reduced to an inequality 
using the fact that the $ n_i $ of the form (\ref{deltani}) do not contribute to the sum (\ref{eps2sum}).  Thus as long as we have
\begin{align}
\sum_{\substack{n_1,\dots,n_M \\ (\sum_i n_i >1 )}}  w_{n_1 \dots n_M} n_i & \le  \alpha_i,
\label{consta}
\end{align}
then we can always choose positive 
\begin{align}
w_{0\dots010\dots0} =  \alpha_i - \sum_{\substack{n_1,\dots,n_M \\ \sum_i n_i >1 }}  w_{n_1 \dots n_M} n_i, 
\end{align}
to satisfy (\ref{epsconst}), where $w_{0\dots010\dots0} $ is the coefficient with $n_i = 1 $ and all other $ n_j = 0 $, $ j \ne i $.

In summary, we must look for coefficients $ w_{n_1 \dots n_M} $ such that (\ref{eps2sum}) is 
satisfied under the constraints of  (\ref{consta}) and (\ref{positivec}). The existence of such a solution can be shown using Farkas' lemma.  Let us begin by formulating our problem in terms of a linear program of the form
\begin{align}
\sum_{0 < k < 2^M} &w_k n_{i}(k) n_{j}(k)  = \alpha_i \alpha_j & (i,j) &\in [0,M]^2 \label{mproof_p1}, \\
\sum_{0 < k < 2^M} &w_k n_{i}(k)  \leq \alpha_i  & 0 \leq i &\leq M\label{mproof_p2}, \\
&w_k \geq 0 & 0 \leq k &< 2^M \label{mproof_p3} .
\end{align}
Here we have relabeled the indices $ n_1 \dots n_M \rightarrow k $ where 
\begin{align}
k = \sum_{j=1}^M 2^{j-1} n_j
\end{align}
is the integer corresponding to the binary number $ n_1 \dots n_M $. The function $ n_j (k) $ returns the $j$th digit of binary representation of integer number $k$. 

This problem can be re-written in canonical form, in matrix notation $A w \leq b$, where $A$ is a matrix of coefficients and $b$ is a vector corresponding to right hand side of the constraints: 
\begin{align}
\sum_{0 < k < 2^M} &w_k n_{i}(k) n_{j}(k) \leq \alpha_i \alpha_j & (i,j) &\in [0,M]^2 \label{primal_1}, \\
-\sum_{0 < k < 2^M} &w_k n_{i}(k) n_{j}(k) \leq -\alpha_i \alpha_j & (i,j) &\in [0,M]^2 \label{primal_2}, \\
\sum_{0 < k < 2^M} &w_k n_{i}(k)  \leq \alpha_i  &0 \leq i &\leq M \label{primal_3},\\
-&w_k \leq 0 & 0 \leq k &< 2^M \label{primal_4}
\end{align}
This representation is semantically equivalent to the representation (\ref{mproof_p1}), (\ref{mproof_p2}), (\ref{mproof_p3}) but has an advantage of having each constraint in the same form. From now on we will refer to this definition as the primal problem.

To prove that primal problem always has a solution we will use Farkas' lemma in the form as stated in Ref. \cite{gale1960theory}, that a solution to $A w \leq b$ exists if and only if an associated dual problem in the form 
\begin{align}
u^T A = 0, \nonumber \\
u \geq 0, \nonumber \\
u^T b < 0,
\end{align}
has no solution. To apply this technique we need to construct the dual problem starting from the primal problem and then show that the dual problem has no solution, or in other words, is infeasible.

Variables in the primal problem are represented by vector $w$, and in the dual problem are represented by the vector $u$.  It is important to note the domains of those variables are not the same as in the dual problem we require $u \geq 0$. From the definition of Farkas' lemma we can deduce that dual problem has as many variables as primal problem had rows and as many equality constraints as primal problem has variables. This indicates that the dual problem must have $2M^2 + M + 2^M$ columns, $u^T A = 0$ creates $2^M$ rows which are equality constraints and $u^T b < 0$ creates a single row that corresponds to a constraint of the form $< 0$ that has only $\alpha_i$ values as coefficients. From this definition we can also see that each row of primal gives a column in the dual, if we group terms we deduce that (\ref{primal_1}) and (\ref{primal_2}) will give us difference of terms with same coefficients, (\ref{primal_3}) will give us a term that is present in $2^M - M$ rows as it vanishes for $M$ rows, and finally, (\ref{primal_4}) will add a distinct variable at the end of each row. We can represent those relevant column groups using the variables $u^{(1)}_{ij}$, $u^{(2)}_{ij}$, $u^{(3)}_{i}$ and $u^{(4)}_{k}$  corresponding to following rows in the primal problem (\ref{primal_1}), (\ref{primal_2}), (\ref{primal_3}) and (\ref{primal_4}) respectively. We also recall that $0 < k < 2^M$.

This gives us enough information to state the dual problem as follows:
\begin{align}
\sum_{\subalign{i,j \\ i \neq j}}^{M} &  n_i(k) n_{j}(k) u^{(1)}_{ij}
- \sum_{\subalign{i,j \\ i \neq j}}^{M} n_{i}(k) n_{j}(k) u^{(2)}_{ij} \nonumber \\
& + \sum_{i}^M n_i(k) u^{(3)}_{i} - u^{(4)}_{k} = 0,  \label{mproof_c1} \\
 \sum_{\subalign{i,j \\ i \neq j}}^{M} &\alpha_i \alpha_j u^{(1)}_{ij} - \sum_{\subalign{i,j \\ i \neq j}}^{M} \alpha_i \alpha_j u^{(2)}_{ij} + \sum_{i}^M \alpha_i u^{(3)}_{i} < 0 . \label{mproof_c3}
\end{align}
In the primal problem, as mentioned earlier, value of $n_{i}(k)$ depends on the value $k$ which is associated to column variable $w_k$. In the dual problem, the value of $k$ is associated to the row number of the problem. Since $n_{i}(k)$ still depends on $k$, each row in the dual problem has $n_{i}(k)$ corresponding to the same binary string.  We can rewrite it to more compact form by adding $u^{(4)}_{k}$ to both sides of (\ref{mproof_c1}). Since $u \geq 0$ it relaxes those constraints from equality constraints into constraints that must be greater or equal to zero. To improve readability, let us also factor out the sum coefficients to give
\begin{align}
\sum_{\subalign{i,j \\ i \neq j}}^{M} n_{i}(k) n_{j}(k) ( u^{(1)}_{ij}
- u^{(2)}_{ij}) + \sum_{i}^M n_{i}(k) u^{(3)}_{i} &\geq 0 ,
& \\
\sum_{\subalign{i,j \\ i \neq j}}^{M} \alpha_i \alpha_j ( u^{(1)}_{ij} - u^{(2)}_{ij}) + \sum_{i}^M \alpha_i u^{(3)}_{i} &< 0 . 
\end{align}
This is a complete dual problem, however we do not need all the constraints to be violated to show dual is infeasible. Showing that one of the 
instances is violated is sufficient to show that the dual problem has no solution.  For convenience we choose the case $k = 2^M - 1$, 
which corresponds to $ n_j = 1 $ for all $ j$:  
\begin{align}
\sum_{\subalign{i,j \\ i \neq j}}^{M} ( u^{(1)}_{ij}
- u^{(2)}_{ij}) + \sum_{i}^M u^{(3)}_{i} &\geq 0, \label{mproof_c4} \\
\sum_{\subalign{i,j \\ i \neq j}}^{M} \alpha_i \alpha_j ( u^{(1)}_{ij} - u^{(2)}_{ij}) + \sum_{i}^M \alpha_i u^{(3)}_{i} &< 0. \label{mproof_c5}
\end{align}
Since both constraints (\ref{mproof_c4}) and (\ref{mproof_c5}) have same value on their right hand side, we can relate them to each other, then we subtract (\ref{mproof_c4}) from all terms
\begin{align}
\sum_{\subalign{i,j \\ i \neq j}}^{M} (\alpha_i \alpha_j - 1) ( u^{(1)}_{ij} - u^{(2)}_{ij}) + \sum_{i}^M (\alpha_i - 1) u^{(3)}_{i} & & \nonumber \\
< \sum_{\subalign{i,j \\ i \neq j}}^{M}(0 - 1)( u^{(1)}_{ij}
- u^{(2)}_{ij}) + \sum_{i}^M (0 - 1) u^{(3)}_{i} &\leq 0. \label{mproof_final}
\end{align}
We can see that (\ref{mproof_final}) cannot be true because $0 \leq \alpha_i \leq 1$, which shows that constraint (\ref{mproof_c1}) with $k = 2^M - 1$ is in conflict with constraint (\ref{mproof_c3}). This shows that the dual problem is not feasible, and therefore by Farkas' lemma, the primal problem is always feasible.

\section{Mean-field theory of the adibatic quantum computing Hamiltonian}

\subsection{Ground state by self-consistent iteration}

In this section we derive a mean-field theory of the Hamiltonian (3) of the main text, using the ensemble Hamiltonians (4) and (5).  Mean-field theory amounts making the substitution
\begin{align}
S_i^z = \langle S_i^z  \rangle + \delta S_i^z
\label{meanfieldapprox}
\end{align}
where $ \delta S_i^z = S_i^z - \langle S_i^z  \rangle $.  The averages $ \langle S_i^z  \rangle $ are unknown at this stage but will be determined later.   Substituting (\ref{meanfieldapprox}) into (3) of the main text, and discarding  second  and higher order terms in $ \delta S_i^z $, we obtain
\begin{align}
H_{\text{MF}}^{(0)} = & \sum_{i=1}^M \left[ - (1-\lambda) S_i^x + \lambda (2M_i + K_i)  S_i^z \right] \nonumber \\
& - \frac{\lambda}{N} \sum_{ij} J_{ij} \langle  S_i^z \rangle \langle  S_j^z \rangle  .
\label{meanfieldham}
\end{align}
where
\begin{align}
M_i = \frac{1}{N} \sum_j J_{ij} \langle S_j^z \rangle  .
\end{align}
The Hamiltonian (\ref{meanfieldham}) can be diagonalized by the transformation
\begin{align}
H_{\text{MF}}^{(0)} = & - \sum_{i=1}^M  \sqrt{ (1-\lambda)^2 + \lambda^2 (2M_i + K_i)^2} \tilde{S}_i^z \nonumber \\
 & - \frac{\lambda}{N} \sum_{ij} J_{ij} \langle  S_i^z \rangle \langle  S_j^z \rangle
\label{diaghammf}
\end{align}
where 
\begin{align}
\tilde{S}_i^z & = \sin \phi_i  S_i^x + \cos \phi_i  S_i^z  
\end{align}
and
\begin{align}
\sin \phi_i & = \frac{1-\lambda}{ \sqrt{ (1-\lambda)^2 + \lambda^2 (2M_i + K_i)^2}} \nonumber \\
\cos \phi_i & = \frac{-\lambda (2M_i + K_i) }{ \sqrt{ (1-\lambda)^2 + \lambda^2 (2M_i + K_i)^2}}  .
\label{sinphicosphi}
\end{align}
The ground state solution of this Hamiltonian takes the general form 
\begin{align}
| \Psi_{\text{MF}}^{(0)} \rangle = \prod_{i=1}^M | 0, \theta_i \rangle_i
\label{groundsupp}
\end{align}
where the state $ | 0, \theta_i \rangle_i $ is defined in (10) of the main text.  This gives the mean-field ansatz of (9) in the main text.  We note that similar mean-field ansatz were used in past works to analyze the $N = 1 $ AQC Hamiltonian ground state \cite{smolin2014classical, shin2014quantum, PhysRevA.94.062106,albash2015reexamining}.

The parameters $ \theta_i $ are found such that the ensembles are maximally polarized, demanding that
\begin{align}
\tilde{S}_i^z | 0, \theta_i \rangle_i = N | 0, \theta_i \rangle_i .
\end{align}
This is satisfied by taking $ \theta_i = \phi_i $, or 
\begin{align}
| 0, \theta_i \rangle_i  = \prod_{n=1}^N \left( \cos  \frac{\phi_i}{2} |0 \rangle_{ni} + \sin  \frac{\phi_i}{2} |1 \rangle_{ni} \right)  .
\label{mfsolution}
\end{align}

Since the angles $ \phi_i $ in  (\ref{sinphicosphi}) involve the unknown expectation values $ \langle S_i^z \rangle $, this still does not constitute a solution.  The find the $ \langle S_i^z \rangle $, we use the solution (\ref{groundsupp}) and (\ref{mfsolution}) to evaluate the expectation value
\begin{align}
\langle S_i^z \rangle = N \cos \phi_i.
\end{align}
Using the expression (\ref{sinphicosphi}) we obtain the self-consistent equation
\begin{align}
x_i = \frac{-\lambda (2 M_i + K_i) }{ \sqrt{ (1-\lambda)^2 + \lambda^2 (2M_i + K_i)^2}} 
\end{align}
where we have used the parametrization $ x_i = \langle S_i^z \rangle/N $. The parameter $ x_i \in [-1,1] $ according to the current definition.  Using the result that the ground state for the qubit and ensemble Hamiltonians are equivalent, we may further parameterize 
\begin{align}
z_i = \sigma_i x_i = \frac{\langle S_i^z \rangle \sigma_i}{N }
\label{zparam}
\end{align}
such that we expect that $ z_i \in [0,1] $. Here $ \sigma_i $ is the ground state configuration for the qubit Hamiltonian $ H_Z $.  The self-consistent equation in terms of $ z_i $ then reads
\begin{align}
z_i = \frac{-\sigma_i \lambda (2 M_i + K_i) }{ \sqrt{ (1-\lambda)^2 + \lambda^2 (2M_i + K_i)^2}} 
\end{align}
where
\begin{align}
M_i = \sum_j J_{ij} \sigma_j z_j  .
\end{align}
The ground state energy according to (\ref{diaghammf}) is then 
\begin{align}
E_{\text{MF}}^{(0)} = & -N  \sum_{i=1}^M  \sqrt{ (1-\lambda)^2 + \lambda^2 (2M_i + K_i)^2} \nonumber \\
 & - N \lambda \sum_{ij} J_{ij} \sigma_i  \sigma_j z_i z_j  .
\end{align}

\subsection{Ground state by optimization}

An equivalent procedure to obtain the ground state is simply to treat (\ref{groundsupp}) as an ansatz wavefunction and optimize for the parameters $ \theta_i $.  Evaluating the expectation value with respect to the Hamiltonian  (3) of the main text, using the ensemble Hamiltonians (4) and (5) yields
\begin{align}
E_{\text{MF}}^{(0)} & = \langle \Psi_{\text{MF}}^{(0)}  | H | \Psi_{\text{MF}}^{(0)} \rangle  \nonumber \\
& = - N ( 1- \lambda ) \sum_{i=1}^M \sin \theta_i \nonumber \\
& + N \lambda \left( \sum_{ij} J_{ij} \cos \theta_i \cos \theta_j + \sum_i K_i \cos \theta_i \right) .
\label{groundopt}
\end{align}
The equivalent parametrization to (\ref{zparam}) corresponds to 
\begin{align}
\cos \theta_i & = z_i \sigma_i \nonumber \\
\sin \theta_i  & = \sqrt{1-z_i^2} ,
\label{thetatozparam}
\end{align}
where $ \sigma_i $ are the ground state spin configurations for the qubit Hamiltonian, and we used the fact that 
\begin{align}
\langle \Psi_{\text{MF}}^{(0)}  | S_i^z | \Psi_{\text{MF}}^{(0)} \rangle  = N  \cos \theta_i .
\end{align}
The ground state energy can then be written as
\begin{align}
E_{\text{MF}}^{(0)} & = - N ( 1- \lambda ) \sum_{i=1}^M \sqrt{1-z_i^2} \nonumber \\
& + N \lambda \left( \sum_{ij} J_{ij} \sigma_i \sigma_j z_i z_j + \sum_i K_i \sigma_i z_i   \right) .
\label{groundzparam}
\end{align}
This expression is optimized for $ z_i \in [0,1] $

This yields the same results as the self-consistent procedure in the previous section.  The self-consistent solution tends to numerically give faster results and hence this is used for our computations.

\subsection{First excited state and gap energy}

To obtain the gap, we require also an estimate of the first excited state.  To deduce the form of this, first let us consider several limiting cases. 

In the limit $ \lambda = 1 $, for parameters such that (8) in the main text is the gap energy (i.e. $ \delta < \Delta $), the first excited state takes the form
\begin{align}
| 1, \frac{(\sigma_k+1)\pi}{2} \rangle_k \prod_{i\ne k } |0, \frac{(\sigma_i+1)\pi}{2} \rangle_i 
\end{align}
where $ k $ is the minimal value found in (8) in the main text. Here we defined the spin coherent states with one of the spins flipped as
\begin{align}
| 1, \theta \rangle  = & \tilde{S}_i^x | 0, \theta \rangle \nonumber \\
= & \frac{1}{\sqrt{N}} \sum_{k=1}^N 
\left( \sin \theta |0 \rangle_k  - \cos \theta  |1\rangle_k \right)  \nonumber \\
& \otimes \prod_{n \ne k } \left( \cos \theta |0 \rangle_i + \sin \theta |1 \rangle_i \right)  .
\label{oneflipstate}
\end{align}
where $ \tilde{S}_i^x = - \cos \theta S^x_i + \sin \theta S^z_i $. The flipped spin is a symmetric superposition across the whole ensemble in (\ref{oneflipstate}). We work in the symmetric subspace because the adiabatic quantum computing Hamiltonian (4) and (5) in the main text only involves symmetric operators.  

In the reverse limit of $ \lambda = 0 $, the first excited state is the state with a single spin flip on one of the ensembles
\begin{align}
| 1,\frac{\pi}{2} \rangle_j \prod_{i\ne j} \left(  | 0, \frac{\pi}{2} \rangle_i  \right) .
\label{lambda0limit}
\end{align}
The first excited state for (\ref{lambda0limit}) is $ M $-fold degenerate, the ensemble with the flipped spin can be any one of $ j \in [1,M] $.  

For small but non-zero $ \lambda > 0 $, the $ H_Z $ will break the degeneracy of the ensemble with the flipped spin.  The lowest energy state will be a superposition of the terms of the form (\ref{lambda0limit}).  
This suggests that we use a mean-field ansatz for the first excited state as
\begin{align}
| \Psi_{\text{MF}}^{(1)} \rangle = \sum_{k=1}^M \psi_k | 1, \theta_k \rangle_k \prod_{i\ne k} | 0, \theta_i \rangle_i  ,
\label{firstexcitedapp}
\end{align}
which gives the expression in the main text. 

We now describe how to find the parameters in (\ref{firstexcitedapp}).  For the parameters $ \theta_i $, as the state is a perturbation of the ground state (\ref{groundsupp}), we simply use the same parameters found in the self-consistent calculation of the ground state.  The $ \psi_k $ can be found by constructing a matrix in the basis
\begin{align}
|\psi_k \rangle = | 1, \theta_k \rangle_k \prod_{i\ne k} | 0, \theta_i \rangle_i
\end{align}
which form an orthogonal set of basis states.  The diagonal matrix elements of the $ M \times M $ matrix can be computed to be
\begin{align}
\langle \psi_k | H | \psi_k \rangle =&  E_{\text{MF}}^{(0)} + 2 (1-\lambda) \sin \theta_k - 2 \lambda K_k \cos \theta_k   \nonumber \\
& - 4 \lambda \sum_{i \ne k} J_{ik}  \cos \theta_i \cos \theta_k 
\end{align}
where  $ E_{\text{MF}}^{(0)} $ is the expression for the ground state energy (\ref{groundopt}),  and we used the fact that $ J_{ij} = J_{ji} $ and $ J_{ii} = 0 $.  In terms of the parametrization (\ref{thetatozparam}), the diagonal terms are
\begin{align}
\langle \psi_k | H | \psi_k \rangle =&  E_{\text{MF}}^{(0)} + 2 (1-\lambda) \sqrt{1-z_k^2} 
- 2 \lambda K_k \sigma_k z_k   \nonumber \\
& - 4 \lambda \sum_{i \ne k} J_{ik} \sigma_i \sigma_k  z_i z_k
\label{diagz}
\end{align}
and the expression for the ground state (\ref{groundzparam}) is used for $ E_{\text{MF}}^{(0)}  $.  The 
off-diagonal terms are 
\begin{align}
\langle \psi_{k'} | H | \psi_k \rangle =& 2 \lambda J_{k k'} \sin \theta_k \sin \theta_{k'} . 
\end{align}
In terms of the parametrization (\ref{thetatozparam}), this can be written
\begin{align}
\langle \psi_{k'} | H | \psi_k \rangle =& 2 \lambda J_{k k'} \sqrt{1-z_k^2} \sqrt{1-z_{k'}^2}  . 
\label{offdiagz}
\end{align}

To calculate the first excited state energy, we diagonalize the matrix defined by (\ref{diagz}) and (\ref{offdiagz}) and take the smallest eigenvalue.  Equivalently, the gap can be directly found by subtracting the ground state energy $  E_{\text{MF}}^{(0)} $ from  (\ref{diagz}), diagonalizing the matrix, and taking the minimum eigenvalue.

\section{Minimum gap for the ferromagnetic case}

Here we examine the minimum gap for the ferromagnetic Hamiltonian in the mean-field limit for various problem sizes $ M $ and compare the behavior 
for the original qubit case $ N = 1 $. 

We first discuss some elementary properties of the ferromagnetic Hamiltonian. 
The ferromagnetic Hamiltonian with bias field $ K $ is defined as
\begin{align}
J_{ij} & = -1 ( 1- \delta_{ij}) \nonumber \\
K_i & = K .
\label{ferromagham}
\end{align}
The ground state is the state with all spins $ S^z_i = - N $ for $ K > 0 $. The energy landscape corresponds to one global minimum and one local minimum separated by a potential barrier consisting of all the remaining states.   For the ensemble case and in the regime $ N > N_c $, the ground state has the same logical configuration $ S^z_i = - N $ but the first excited state is a single spin flip of the configuration $ S^z_k = - N+2, S^z_i = - N $ $\forall i \ne k $. Due to the symmetry between all the sites $ i $, the first excited state is $M $-fold degenerate.  The crossover occurs for $ N $ such that $ \Delta > \delta $ where in this case 
\begin{align}
\Delta & = 2KNM \nonumber \\
\delta & = 4(M-1) + 2K .  
\end{align}
This gives a critical ensemble size for the ferromagnetic case as
\begin{align}
N_c = \lfloor \frac{2M+ K -2}{KM} + 1  \rfloor .  
\end{align}

\begin{figure}[t]
\centering
\includegraphics[width=\linewidth]{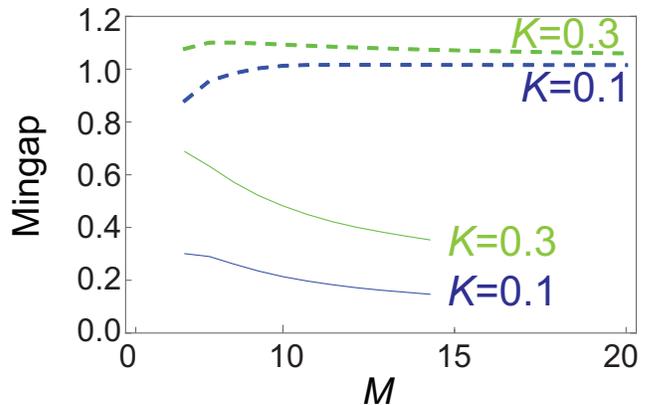} 
\caption{The minimum gap of the ferromagnetic model with $ J_{ij} = -(1-\delta_{ij}) $ and $ K_i = K $ for $ N =1 $ (solid lines) and from mean-field theory ($ N \rightarrow \infty $; dashed lines).   \label{figs2}}
\end{figure}

 Fig. \ref{figs2} shows the minimum gap using the mean-field theory (i.e. $ N \rightarrow \infty $) of the previous section.  For comparison, the 
minimum gap for the $ N = 1 $ case is also calculated for the Hamiltonian (3) using (4) and (5) in the main text by direct diagonalization.  We see that for the qubit case the gap decreases with $ M $, 
but the decay is less rapid for the mean-field case.  The differing behavior can be attributed to the fact that the nature of the first excited state is quite different for the two limits.  For $ N = 1 $ (which by definition is always the $N < N_c $ regime), 
the first excited state in the limit  $ \lambda \rightarrow 1 $ is the other ferromagnetic state $ S^z_i = +N = +1 $. The minimum gap is 
then found by obtaining the $ \lambda $ such that the transverse field produces the smallest gap, which is likely to admix the two ferromagnetic states.  In contrast, the first excited state for the mean-field case is always a logically equivalent state, since it is by definition in the $N > N_c $ regime (since $ N \rightarrow \infty $). The differing nature of the first excited state gives rise to a different scaling of the gap. In this case, 
the ensemble approach is expected to perform well also for large problem sizes $ M $ due to the larger gap in the ensemble case. 

\section{Decoherence}
 
In the main text we discussed the performance of AQC under collective dephasing where all states are restricted to the symmetric subspace, and only total spin operators appear in the master equation.  The collective dephasing model is especially relevant to atomic ensembles, but for other implementations --- such as those using superconducting qubits --- the decoherence will occur for the individual physical qubits.  In this case, the appropriate master equation is instead
\begin{align}
\frac{d \rho}{dt}= & i [\rho,H]
-\Gamma_z \sum_{n=1}^{M} \sum_{k=1}^{N} [\rho  - \sigma^z_{n,k} \rho \sigma^z_{n,k} ] .
\label{individualmaster}
\end{align}
For the collective decoherence model (Eq. (13) in the main text), all states are restricted to the symmetric subspace, since only total spin operators appear in the master equation.  In this case, the dimension of each ensemble is $ N + 1 $, giving a total Hilbert space dimension of $ (N+1)^M $.  For individual qubit dephasing, the state no longer is restricted to the symmetric subspace, hence the full set of $ 2^N $ states must be considered for each ensemble.  The total Hilbert space dimension is $ 2^{NM} $ in this case, which gives a much larger numerical overhead.

\begin{figure}[t]
\centering
\includegraphics[width=\linewidth]{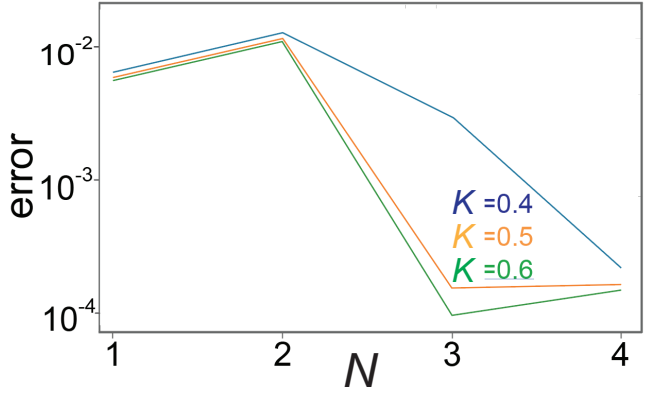} 
\caption{Error probabilities for the ferromagnetic Hamiltonian (\ref{ferromagham}) with $ M = 3 $ under individual qubit dephasing (\ref{individualmaster}).  The decoherence rate is taken as $ \Gamma_z = 10^{-4} $.  \label{figs3}}
\end{figure}

To observe the behavior for the individual dephasing case, we consider the ferromagnetic problem (\ref{ferromagham}) and perform the AQC to evaluate the logical errors, in a similar way to Fig. 4(a) in the main text.  The results are shown in  Fig.  \ref{figs3}. Due to the larger Hilbert space dimension, the largest size that we were able to simulate was for $ N = 4 $.  Within this range, we observe a sharp decrease of errors with increasing $N $, consistent with what is observed in Fig. 4(a) of the main text.

\begin{figure}[t]
\centering
\includegraphics[width=\linewidth]{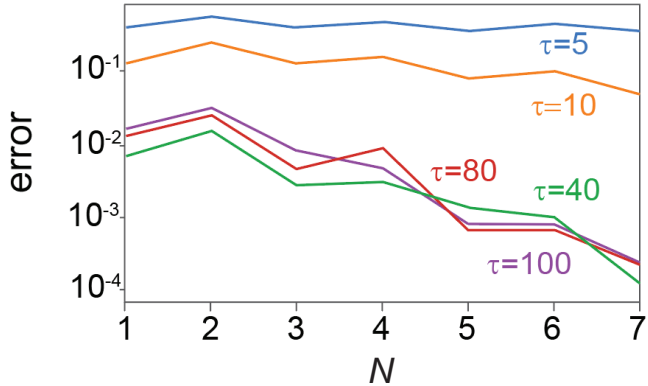} 
\caption{Error versus $N$ for various $\tau$ for a hard Exact Cover instance with $ M=3 $ and $\Gamma=10^{-4}$. The parameters of the Exact Cover instance are $J_{12}=0.5, J_{13}=0, J_{23}=0.5, K_1=-1, K_2=-1, K_3=-1 $.   \label{figs4}}
\end{figure}

In addition to randomly generated problem instances, we also examine the Exact Cover 3 problem under collective $ S^z$-dephasing\cite{farhi2001quantum}, as an illustration of the compatibility of the Hamiltonian (4) in the main text with combinatorial problems.  We choose hard Exact Cover 3 instances, which are defined as problems with a unique assignment of values as solutions, corresponding to a non-degenerate ground state of the problem Hamiltonian. The scaling of the error with $ N $ for various adiabatic sweep times $ \tau $ is shown in   Fig. \ref{figs4} for a typical problem instance. In a similar way to Fig. 4(c), we see again that the error scales favorably with $ N $ as long as the sweep time is large enough to maintain adiabaticity but is within the decoherence window. We have verified that similar results are obtained for other generated instances of Exact Cover.

\section{Ensemble-ensemble entanglement}

\begin{figure}[t]
\centering
\includegraphics[width=\linewidth]{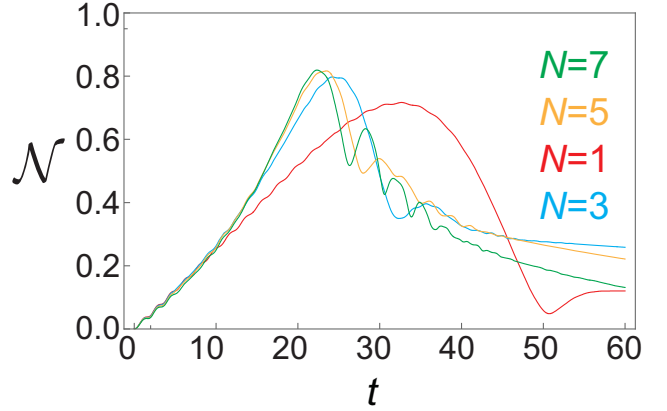} 
\caption{Entanglement between ensembles in an adiabatic evolution.  The logarithmic negativity $ {\cal N } = \log_2 || \rho^{T_2} || $ is calculated for the $ M = 2 $ case for $ J_{ij} = -1 (1- \delta_{ij}) $, $ K_i = 0.1 $, an adiabatic sweep time of $ \tau = 60 $, and dephasing rate of $ \Gamma = 10^{-4} $.    \label{figs1}}
\end{figure}

The mean-field ground state Eq. (9) in the main text is explicitly of the form of a product state, which suggests that there is zero entanglement between the ensembles at all times in the adiabatic evolution.  This is in fact an artifact of the mean-field approximation, and typically there will be entanglement between the ensembles.  In   Fig. \ref{figs1} we show the entanglement between two ensembles as characterized by the logarithmic negativity \cite{vidal2002computable,plenio2005logarithmic} during the adiabatic sweep.  We see that as the ensemble size is increased, the entanglement does not diminish and approaches a common curve. This is consistent with prior studies relating to the robustness of entanglement in such ensembles in the presence of decoherence.  The basic result is that for interaction times of the $ S^z_i S^z_j $ Hamiltonian of the order $ t \sim 1/N $, the entanglement survives robustly in the limit of $ N \rightarrow \infty $ \cite{PhysRevA.88.023609}.  Due to the factor of $ 1/N $ in the ensemble Hamiltonian $ H_Z $, the class of entanglement that is created by the AQC Hamiltonian is expected to be similar.

\section{Experimental implementation}

In this section, we explain in more detail the experimental implementations of our Hamiltonian (4) and (5) in the main text, based on neutral atomic ensembles. 

\subsection{The physical system}

Two physical implementations with neutral atoms are possible, depending upon the operating temperature: thermal atomic ensembles near room temperature and cold atom ensembles near quantum degeneracy.   For thermal atomic ensembles, the atoms are trapped in paraffin-coated glass cells, which preserve the coherence of the internal states of the atoms.  Such systems have been already experimentally studied by numerous groups, notably the Polzik group, see Ref.  \cite{RevModPhys.82.1041} for a review. Entanglement between atomic ensembles each with $ N \sim 10^{12}$ has been achieved \cite{julsgaard2001experimental}, as well as continuous variables teleportation \cite{sherson2006quantum,krauter2013deterministic}. Such atomic ensembles can be scaled up having more glass cells.  For example, in Ref. \cite{pu2017experimental} a system of $ M = 225$ locally addressable atomic ensembles was realized.  Furthermore, entanglement shared between $ M = 25$ such ensembles was realized in Ref. \cite{pu2018experimental}. The logical states are defined by the internal hyperfine ground states which are known to have long coherence times.

Another possibility is using cold atoms trapped on atom chips \cite{reichel2011atom}.  In this case, multiple traps are patterned on the same atom chip, such as that proposed in Ref. \cite{abdelrahman2014coherent}.  In this case, each atomic cloud is trapped magnetically, rather than by a glass cell for the case with thermal ensembles.   The logical states that are used for the case of  ${}^{87}\text{Rb}$ are the  $ F=1,m_F=-1 $ and $F=2,m_F=1 $ ground states \cite{bohi2009coherent, riedel2010atom}. These states are typically used firstly because they are both magnetically trapped states.  Secondly, this pair of states responds in the same way in the presence of magnetic field fluctuations.  Hence while the wires of the atom chip produce magnetic fields with some noise due to current fluctuations, the effect of this is not detrimental to the coherence of the atoms. The magnetically trapped atoms are laser cooled and initialized to the $ F=1,m_F=-1 $ state.  One advantage of the cold atom system is that long coherence can be achieved.  For ${}^{87}\text{Rb}$ atoms at a temperature of 175 nK coherence times of nearly 1 minute were reported in Ref. \cite{deutsch2010spin}.  The long coherence times were attributed to the Identical Spin Rotation Effect (ISRE), where the spins spontaneously rephase in an analogous way to spin echo methods.  Entanglement between two Bose-Einstein condensates (BECs) in separate traps has not been realized experimentally at the time of writing. However, entanglement between two different spatial regions of the same BEC have been detected \cite{kunkel2017,lange2017,fadel2017}.  Numerous theoretical schemes for entanglement between BECs have been proposed \cite{pyrkov2013entanglement,hussain2014geometric,pettersson2017light,abdelrahman2014coherent,treutlein2006microwave,jing2018split,idlas2016entanglement,ortiz2018adiabatic}.  This has been suggested for use in various quantum information protocols such as teleportation \cite{pyrkov2014quantum,pyrkov2014full}, and quantum computation \cite{byrnes2012macroscopic,byrnes2015macroscopic,semenenko2016implementing}.

\subsection{Spinor quantum computing}

Much of the past work relating to the experimental implementation of the error suppressed AQC that is described in the main text has been done in the context of Spinor Quantum Computing (SQC) \cite{byrnes2012macroscopic,byrnes2015macroscopic,semenenko2016implementing,pyrkov2014quantum,pyrkov2014full,byrnes2020quantum}.  We give a brief description of the SQC scheme to relate these past works to our current context. 

Let us first explain how to store and manipulate quantum information in the SQC scheme. The quantum information of a  standard qubit 
\begin{align}
   \alpha \vert 0\rangle +\beta \vert 1\rangle 
\end{align} 
is encoded on ensembles of two-level systems, using a simple duplication strategy
\begin{equation}
| \alpha, \beta \rangle \rangle  =\prod_{n=1}^{N}( \alpha  \vert 0\rangle_{n}+ \beta  \vert 1\rangle_{n}) ,\label{a}
\end{equation}
which are simply spin coherent states \cite{byrnes2015macroscopic}. 
Here $ \vert 0\rangle_{n} $ and $ \vert 1\rangle_{n} $ are logical states of the $ n $th particles in the ensemble and $ \alpha $ and $ \beta $ are complex numbers which satisfy $ \vert \alpha\vert^{2}+ \vert \beta\vert^{2}=1 $.  
The ``double ket'' notation above is mathematically no different to a regular ket, but gives the reminder that this is a macroscopic state.  The encoding makes the state non-linear regarding the coefficients $ \alpha, \beta $ for $ N >1 $
\begin{align}
    | \alpha, \beta \rangle \rangle  \ne \alpha | 1, 0 \rangle \rangle + \beta  | 0,1 \rangle \rangle  .
\end{align}
When dealing with cold atom ensembles below the BEC critical temperature, a spin coherent state is more appropriately written as
\begin{equation}
\vert \alpha ,\beta \rangle \rangle =\frac{1}{\sqrt{N!}}(\alpha a^{\dagger}+\beta b^{\dagger})^{N}\vert 0\rangle,\label{b}
\end{equation}
The $ a $ and $ b $ are bosonic annihilation operators which satisfy the commutation relations $ [a,a^{\dagger}]=[b, b^{\dagger}]=1 $ and  correspond to the two logical states that store the quantum information. $ N $ is the number of atoms in either case, which is a conserved number for coherent operations.   

The main idea of SQC is that such an encoding can be used instead of standard qubits and manipulated in an analogous way. The benefit of this is primarily duplication, which leads to logical error suppression, as demonstrated in the main text. Typically the initial state is prepared in the product state where all the ensembles are in the state $ | 1, 0 \rangle \rangle $.  Then Hamiltonians only consisting of the total spin operators $ S_i^{x,y,z} $ are applied in sequence. This, as we explain later, ensures that the experimental resources remain tractable.   Finally, measurements are made by reading off the particle numbers in the Fock basis  
\begin{align}
|k \rangle = \frac{ (a^\dagger)^{k} (b^\dagger)^{N-k}}{\sqrt{k! (N-k)! }} | 0 \rangle .
\label{fockstates}
\end{align}
for each ensemble.  From the measurement readout, a suitable encoding is performed in order to relate the outcome to the original problem. In the case of this work, a majority vote is done such that $ k< N/2 $ corresponds to a logical 0 and $ k > N/2$ is a logical 1.  

Let us consider some elementary examples to show the behavior of various operations under the SQC mapping. For single logical qubits based on ensembles, this is done by applying the Hamiltonians of the total spin operators
\begin{align}
    S^{x,y,z} = \sum_{n=1}^N \sigma_n^{x,y,z}
\end{align}
This produces rotations of the encoded states as
\begin{align}
& e^{-i \theta S^x/2 }   | \alpha, \beta \rangle \rangle \nonumber \\
& = | \alpha \cos (\theta/2)  -i \beta \sin (\theta/2) , \beta \cos (\theta/2)  -i \alpha \sin (\theta/2)   \rangle \rangle \nonumber, \\
& e^{-i \theta S^y/2 }  | \alpha, \beta \rangle \rangle \nonumber \\
& = |\alpha  \cos (\theta/2) - \beta \sin (\theta/2) , \beta  \cos (\theta/2) + \alpha \sin (\theta/2)  \rangle \rangle \nonumber, \\
&  e^{-i \theta S^z/2 } | \alpha, \beta \rangle \rangle \nonumber \\
& = |e^{-i \theta/2} \alpha , e^{i \theta/2} \beta \rangle \rangle  .
\label{spinrotationsxyz}
\end{align}
in the same way as regular qubits.  For the case of BECs, the spin operators are instead written 
\begin{align}
S^{x}&=a^{\dagger}b+a b^{\dagger}, \nonumber\\
S^{y}&=i( b^{\dagger} a -a^{\dagger}b),\nonumber\\
S^{z}&=a^{\dagger}a- b^{\dagger}b,
\end{align}
The rotations (\ref{spinrotationsxyz}) are the same for the BEC case.  

We note that despite the rather different appearance of the ensemble and BEC cases, they are mathematically equivalent as long as the applied operations are symmetric with respect to particle interchange.  In the BEC case, the spin coherent states (\ref{b}) can be expanded in terms of Fock states (\ref{fockstates}) which span a $ N +1$ dimensional Hilbert space.  In contrast, in the ensemble case the dimension is $ 2^N$. If the initial state is symmetric under particle interchange, which is the case for spin coherent states (\ref{a}), and all the operations are also symmetric under interchange, then the states are restricted to the symmetric subspace.  This reduces the effective dimension of the ensemble from $ 2^N$ to  $ N +1$ \cite{byrnes2020quantum}.  

For two qubit gates, the types of states that are produced with ensembles are analogous, but do not have the perfect equivalence as single qubits.  A typical type of interaction that might be present between ensembles is the $ H = S^z_i S^z_j $ interaction, where $ i,j $  label two ensembles. For the qubit case, the analogous Hamiltonian is $ H = \sigma^z_i \sigma^z_j $ and can produce an entangled state 
\begin{align}
& e^{-i \sigma^z_i \sigma^z_j \tau}  \left(\frac{|0 \rangle+ |1 \rangle}{\sqrt{2}} \right) \left(\frac{|0 \rangle+ |1 \rangle}{\sqrt{2}} \right)  \nonumber \\
& = \left( \frac{e^{i \tau} | 0 \rangle + e^{-i \tau} | 1 \rangle}{\sqrt{2}} \right) | 0 \rangle + 
\left( \frac{ e^{-i \tau} | 0 \rangle + e^{i \tau} | 1 \rangle }{\sqrt{2}} \right) | 1 \rangle   .
\label{qubitszsz}
\end{align}
For a time $ \tau = \pi/4 $, the above state becomes a maximally entangled state and  can be used as the basis of a CNOT gate in the appropriate basis. 

For ensembles, the $ H =S^z_i S^z_j $ interaction produces the state 
\begin{align}
& e^{-i S^z_i S^z_j \tau} | \frac{1}{\sqrt{2}}, \frac{1}{\sqrt{2}} \rangle \rangle_i 
| \frac{1}{\sqrt{2}}, \frac{1}{\sqrt{2}} \rangle \rangle_j  \nonumber \\
& = \frac{1}{\sqrt{2^N}} \sum_k  \sqrt{N \choose k} | \frac{e^{i(N-2k)\tau}}{\sqrt{2}} , \frac{e^{-i(N-2k)\tau}}{\sqrt{2}} \rangle \rangle_i | k \rangle_j \nonumber \\
& = \frac{1}{\sqrt{2^N}} \Big[ 
| \frac{e^{i N \tau}}{\sqrt{2}} , \frac{e^{-i N\tau}}{\sqrt{2}} \rangle \rangle_i | 0 \rangle_j 
+ \sqrt{N} | \frac{e^{i (N-2) \tau}}{\sqrt{2}} , \frac{e^{-i (N-2) \tau}}{\sqrt{2}} \rangle \rangle_i | 1 \rangle_j  \nonumber \\
& \dots  +  \sqrt{N \choose N/2}   | \frac{1}{\sqrt{2}} , \frac{1}{\sqrt{2}} \rangle \rangle_i | N/2  \rangle_j
+ \dots +
| \frac{e^{-i N \tau}}{\sqrt{2}} , \frac{e^{i N\tau}}{\sqrt{2}} \rangle \rangle_i | N \rangle_j  \Big] .
\label{twobecszsz}
\end{align}
Comparing (\ref{qubitszsz}) and (\ref{twobecszsz}), we see that a similar type of state is produced where the first ensemble is rotated by an angle $ 2 (N - 2k) \tau $ around the $ S^z $ axis, for a Fock state $ |k \rangle $ on the second ensemble.  The types of correlations are similar for both the qubit and ensemble cases, but there are also differences.  The first most obvious difference is that the ensemble consists of $ N +1 $ terms, whereas the qubit version only has two terms. Thus although the same type of correlations are present to the qubit entangled state, it is a higher dimensional generalization.  The other difference is the presence of the binomial factor which weight the terms. The binomial function approximately follows a Gaussian distribution with a standard deviation $ \sim \sqrt{N} $.  Thus the most important terms are in the range $k \in [N/2-\sqrt{N}, N/2+\sqrt{N}] $.  The task in the SQC mapping is then to ensure that the ensemble mapping still produces the relevant logical operations that is desired in the qubit formulation.  In the case of AQC for this work, this is ensured by the results as shown in the Appendix. \ref{sec:equivall}.  

Using a sequence of single and two ensemble Hamiltonians, it is possible to produce a large range of effective Hamiltonians.  A well-known quantum information theorem states that if it is possible to perform an operation with Hamiltonians $ H_i $ and $ H_j $, then it is also possible to perform the operation corresponding to $ H_k = i [H_i,H_j] $ \cite{lloyd1995almost}. The combination of single and two ensemble Hamiltonians  may thus  be combined to form more complex effective Hamiltonians.  Using this theorem one can show that for a $ M $ ensemble system, it is possible to produce any Hamiltonian of the form
\begin{align}
\label{prodspinham}
H_{\text{eff}} \propto \prod_{i=1}^M S_i^{q(i)} .
\end{align}
where $ q(i) \in \{ 0, x, y, z \} $ and $ S^0 \equiv I $.  An arbitrary sum of such Hamiltonians may also be produced by performing a Trotter expansion  \cite{lloyd1995almost}. In general higher order operators can also be constructed (e.g. $ (S^q)^l$ with $ l\ge 2 $). Eq. (\ref{prodspinham}) is the analogous result to universality for gate-based qubit quantum computation, since any Hamiltonian of the same form as that for qubits can be constructed using one and two-ensemble gates.  

The one and two ensemble Hamiltonians $ H_i = \bm{n} \cdot \bm{S}_i $ and  $ H_{ij} = S^z_i S^z_j $ can therefore be used together to perform various quantum information processing tasks.  What should result after the application of the sequence of gates is an encoded version of the qubit information, where the spin readout gives the same logical information as the original qubit version. Currently, there is no unique mapping procedure from qubit quantum computing to SQC, and the particular scheme that is chosen depends upon the particular algorithm that is in question \cite{semenenko2016implementing,pyrkov2014quantum,pyrkov2014full}. Considerations that should be taken into account include the sensitivity of the generated states to decoherence and the complexity of the operations involved. Typically one assumes that the individual qubits that comprise the ensemble cannot be manipulated individually.

\subsection{Implementing one and two ensemble operations}

The single ensemble $ S^x_i $ Hamiltonian is produced using microwave and/or radio frequency transitions. The Hamiltonian (5) in the main text has the same amplitude for all the ensembles and hence there is no need for any local addressing of each ensemble for this term.  This is fortunate  because microwave and radio frequency radiation is less easily directed than optical radiation and would affect all the atomic ensembles/BEC that are in the same vicinity.  For the case of $ ^{87} \text{Rb}$ using the clock states $ F=1, m_F = -1$ and $ F=2, m_F = 1$, a two-photon transition is used to create the effective $ S^x $ operation, and has been demonstrated experimentally in numerous works \cite{bohi2009coherent, treutlein2006microwave, riedel2010atom}. 

For the single ensemble $ S^z_i $ Hamiltonian, single ensemble addressability is required due to the coefficient $ K_i$ in the Hamiltonian (4) of the main text. For the $ ^{87} \text{Rb}$ example, the first order Zeeman shift of the clock states are in the same direction, hence it is not possible to apply local magnetic fields to induce the $ S^z $ Hamiltonian. In Ref. \cite{bohi2009coherent}, an alternative approach based on state-dependent potentials from microwave waveguides was used to shift the energy of the $ F = 1, m_F = -1 $ state.  The process is analogous to the ac Stark shift, but at microwave frequencies.  A second-order transition to the $ F = 2, m_F = -1 $ state shift the energy of the  $ F = 1, m_F = -1 $ state.  The other clock state $ F = 2, m_F = 1 $ remains unaffected since it is off-resonant.  Alternatively, the ac Stark shift at optical frequencies could be used to to shift one of the clock states, taking advantage of forbidden transitions \cite{abdelrahman2014coherent}. 

Finally, we also need to generate the $ S^z_i S^z_j $ interactions weighted by the corresponding $ J_{ij} $ in (4) of the main text. There are two main ways that this type of interaction can be generated, either by optically mediation or interaction-based methods.  Optically mediated methods have been discussed in Refs. \cite{pyrkov2013entanglement,rosseau2014entanglement,abdelrahman2014coherent,hussain2014geometric}.  In Refs. \cite{pyrkov2013entanglement,rosseau2014entanglement,abdelrahman2014coherent}, the common optical mode in a cavity QED Hamiltonian was adiabatically eliminated to show that an effective interaction of the form 
\begin{align}
    H_{\text{eff}} \propto 2  S^z_i S^z_j + ( S^z_i)^2 + ( S^z_j)^2  
    \label{effham1}
\end{align}
was generated.  This corresponds to the desired interactions, in addition to local squeezing terms. The same effective Hamiltonian was derived using a different approach based on geometric phases in Ref. \cite{hussain2014geometric}.  In this case the starting Hamiltonian was a quantum non-demolition Hamiltonian and coherent light illuminates the atomic ensembles such that different phases are picked up by different spins of the atoms.  Such optical methods are convenient since the interaction can be produced between arbitrary pairs of ensembles, by directing the common optical mode between them, and the parameters of each can be suitably controlled. For example, in the case of the approach of Refs.  \cite{pyrkov2013entanglement,rosseau2014entanglement,abdelrahman2014coherent}, the detuning  is a free parameter that can be changed to realize various couplings of $ J_{ij} $.   For the geometric phase approach of Ref. \cite{hussain2014geometric}, the trajectory of the geometric path in phase space determines the coefficient of (\ref{effham1}), which is controllable.

For an interaction-based approach with BECs, one may use state-dependent potentials and interactions such as that proposed in Ref. \cite{treutlein2006microwave}.  In this scheme, two BECs are brought close together in a double-well trap.  Then state-dependent potentials are turned on such that only the atoms in one of the levels, say the $ F = 2, m_F = +1 $ overlap spatially.  These atoms possess a non-linear interaction, which corresponds to a $ S^z_i S^z_j$ interaction. In this case the amount of interaction is controlled by the timing of the application of the state-dependent potentials.  This method relies on physically bringing together the BECs, hence may be more suited to creating interactions between nearest-neighbor BECs.  

We note, for the sake of simplicity in this work we did not consider any on-site interaction terms of the type $ (S^z_i)^2 $. In fact such on-site terms are expected to further reduce the logical errors  \cite{boixo2014evidence,vinci2016nested}.  Interestingly, many of the methods for producing ensemble-ensemble interactions also produce local squeezing as a by-product as seen in (\ref{effham1}).  This is also true for other entangled state generation schemes such as the split-squeezing methods of Ref. \cite{jing2018split}.  In order to match the Hamiltonian (4) of the main text, such intra-ensemble terms can be removed by application of local squeezing terms to antisqueeze the ensembles locally. Such local squeezing has been realized in works such as \cite{riedel2010atom}. The sign of the squeezing can be controlled by the detuning of the second order transition.

\subsection{Experimental resources}

One of the primary assumptions that was made in considering the Hamiltonian (4) and (5) of the main text is the lack of control over individual qubit manipulations. This can be seen in Hamiltonian (4) and (5) as they are entirely composed of the collective spin operators $ S_i^{x,z} $.  
Removing the necessity of individual qubit control in the ensemble allows for a massive level of parallelization of the gate operations in the scheme. For example, in atomic ensembles the duplication number  for cold atoms  is $ N \sim 10^3$ \cite{riedel2010atom, fadel2017} and for thermal ensembles is $ N\sim 10^{12}$ \cite{julsgaard2001experimental}. Despite the very large number of physical qubits, the experimental resources to control do not increase in proportion with $ N $.  For example, in order to implement the $ S^x $ interaction, microwave and radio frequency radiation is applied to the entire ensemble together.  The same operation is performed whether $ N =1 $ or $ N = 10^{12}$ since the atoms evolve in parallel.  We may compare this to related error suppressing approaches where the experimental resources typically increase with $ N $.  We note that this is true for all the operations, including the $ S^z_i S^z_j $ interactions.  In fact, due to the factor of $ 1/N $ in (4) of the main text, shorter entangling pulses would need to be applied as $ N $ is scaled up.  This factor originates from collective enhancement, since the spin operators $ O(S^z_i) \sim N $ and the equivalent entanglement requires interaction times that are $ N $ times smaller \cite{PhysRevA.88.023609}.


\bibliographystyle{apsrev}
\bibliography{references_Naeimeh}

\end{document}